\documentclass[twocolumn,
amsmath,amssymb,aps,prx]{revtex4-1}

\usepackage{graphicx}
\usepackage{dcolumn}
\usepackage{bm}
\usepackage{amssymb,amsmath}
\usepackage{upgreek}

\usepackage[colorlinks,urlcolor=blue,citecolor=blue,linkcolor=blue]{hyperref}

\usepackage{comment}

\renewcommand{\k}{{\bf k}}
\newcommand{\p}{{\bf p}}

\newcommand{\q}{{\bf q}}
\newcommand{\Q}{{\bf Q}}

\renewcommand{\r}{{\bf r}}
\newcommand{\0}{{\bf 0}}

\newcommand{\bra}[1]{\left\langle{#1}\right|}
\newcommand{\ket}[1]{\left|{#1}\right>}
\newcommand{\braket}[2]{\left<{#1}\middle|{#2}\right>}

\newcommand{\eb}{E_{\rm B}}
\newcommand{\eg}{E_{g}}

\newcommand{\nn}{\nonumber}
\newcommand{\beq}{\begin{equation}}
\newcommand{\eeq}{\end{equation}}

\newcommand{\sch}{Schr{\"o}dinger }

\newcommand{\ok}{\bar{\omega}_\k}

\usepackage[usenames,dvipsnames]{color}
\newcommand{\sout}[1]{}

\newcommand{\mmp}[1]{{\color{black}#1}}
\newcommand{\jfl}[1]{{\color{black}#1}}

\newcommand{\Qe}{\mathbf{Q}_e}
\newcommand{\Qh}{\mathbf{Q}_h}
\newcommand{\kP}{\mathbf{k}^\prime}

\newcommand{\diff}{\mathop{}\!\mathrm{d}}

\begin{document}

\title{Microscopic description of exciton-polaritons in microcavities}

\author{Jesper Levinsen}
\affiliation{School of Physics and Astronomy, Monash University, Victoria 3800, Australia}
\affiliation{ARC Centre of Excellence in Future Low-Energy Electronics Technologies, Monash University, Victoria 3800, Australia}

\author{Guangyao Li}
\affiliation{School of Physics and Astronomy, Monash University, Victoria 3800, Australia}
\affiliation{ARC Centre of Excellence in Future Low-Energy Electronics Technologies, Monash University, Victoria 3800, Australia}

\author{Meera M.~Parish}
\affiliation{School of Physics and Astronomy, Monash University, Victoria 3800, Australia}
\affiliation{ARC Centre of Excellence in Future Low-Energy Electronics Technologies, Monash University, Victoria 3800, Australia}

\date{\today}

\begin{abstract}
  We investigate the microscopic description of exciton-polaritons that involves electrons, holes and photons within a two-dimensional microcavity. We show that in order to recover the simplified exciton-photon model that is typically used to describe polaritons, one must correctly define the exciton-photon detuning and exciton-photon (Rabi) coupling in terms of the bare microscopic parameters.  For the case of unscreened Coulomb interactions, we find that the exciton-photon detuning is strongly shifted from its bare value in a manner akin to renormalization in quantum electrodynamics.  Within the renormalized theory, we exactly solve the problem of a single exciton-polariton for the first time and obtain the full spectral response of the microcavity. In particular, we find that the electron-hole wave function of the polariton can be significantly modified by the very strong Rabi couplings achieved in current experiments.  \mmp{Our microscopic approach furthermore allows us to obtain the effective interaction between identical polaritons for any light-matter coupling.  Crucially, we show that the standard treatment of polariton-polariton interactions in the very strong coupling regime is incorrect, since it neglects the light-induced modification of the exciton size and thus greatly overestimates the effect of Pauli exclusion on the Rabi coupling, i.e., the saturation of exciton oscillator strength.}  Our findings \jfl{thus provide the foundations} for understanding and characterizing exciton-polariton systems across the whole range of polariton densities.  
\end{abstract}

\maketitle

\section{Introduction}
A strong light-matter coupling is routinely achieved in experiment by embedding a semiconductor in an optical microcavity \cite{Microcavities}. When the coupling strength exceeds the energy scale associated with losses in the system, one can create hybrid light-matter quasiparticles called polaritons, which are superpositions of excitons and cavity photons~\cite{Keeling2007,Deng2010,Carusotto2013}.  Such exciton-polaritons have been successfully described using a simple model of two coupled oscillators, where the exciton is treated as a rigid, point-like boson. This simple picture underpins the multitude of mean-field theories used to model the coherent many-body states of polaritons observed in experiment, such as Bose-Einstein condensation and superfluidity~\cite{Kasprzak2006, Balili1007, Utsunomiya2008, KohnlePRL2011, Roumpos2010, RobPRL14}.  However, with advances in device fabrication leading to cleaner samples, higher quality cavities and stronger light-matter coupling, experiments are now entering a regime where the composite nature of the exciton plays an important role.

Most notably, the structure of the exciton bound state determines the strength of the polariton-polariton interactions, which are currently a topic of major interest since they impact the many-body physics of polaritons, as well as the possibility of engineering correlations between photons~\cite{Volz2017,Delteil2018}.  In the absence of light, the low-momentum scattering of excitons can be theoretically estimated by considering exchange processes involving electrons and holes~\cite{Ciuti1998,Tassone1999}. However, for the case of polariton-polariton interactions, the exchange processes are complicated by the coupling to photons, and there are currently conflicting theoretical results in the literature~\cite{Tassone1999,Rochat2000,MCombescot2007,Xue2016}.  Moreover, none of these previous works properly include light-induced modifications of the exciton wave function, which can be significant at strong light-matter coupling~\cite{Khurgin2001} and which are crucial for determining the polariton-polariton interaction strength, as we show here.

There is also the prospect of achieving strong light-matter coupling in a greater range of systems, such as atomically thin materials~\cite{Mak2016,Low2016}. In particular, polaritons have recently been realized in transition metal dichalcogenides~\cite{Liu2014,Dufferwiel2015}, where it is possible to electrostatically gate the system and create correlated light-matter states involving an electron gas~\cite{Sidler2016,Knuppel2019}.
Moreover, by increasing the photon intensity, one can access the high-excitation regime of the microcavity~\cite{Horikiri2017}, where the exciton Bohr radius becomes comparable to or exceeds the mean separation between polaritons~\cite{Kamide2010,Byrnes2010,Kamide2011,Yamaguchi2012,Yamaguchi2013}.  Both these scenarios further underline the need for a microscopic description that goes beyond the simple exciton-photon model.

In this paper, we exactly solve the problem of a single exciton-polariton within a low-energy microscopic model of electrons, holes and photons in a two-dimensional (2D) microcavity. In contrast to previous variational approaches to the problem~\cite{Khurgin2001,Zhang2013,Yang2015}, we capture all the exciton bound states and the unbound electron-hole continuum, which are important for describing the regime of very strong light-matter coupling. Furthermore, we find that the cavity photon frequency is renormalized from its bare value by an amount that is set by the ultraviolet (UV) cutoff, since the photon couples to electron-hole pairs at arbitrarily high energies in this model. Physically, this is because the microscopic model includes all electron-hole transitions, as well as the excitonic resonances, that determine the dielectric function of the microcavity.  \jfl{While such behavior has also been observed in classical theories of the dielectric function~\cite{Averkiev2007}, this crucial point has apparently been overlooked by all previous quantum-mechanical treatments of the electron-hole-photon model.}

Within the renormalized microscopic model, we can formally recover the simple exciton-photon model in the limit where the exciton binding energy is large compared with the light-matter (Rabi) coupling. However, for larger Rabi coupling, we find that the exciton wave function is significantly modified, consistent with recent experimental measurements of the exciton radius in the upper and lower polariton states~\cite{Brodbeck2017}. Moreover, we find that the upper polariton becomes strongly hybridized with the electron-hole continuum and thus cannot be described within a simple two-level model in this regime.  Our microscopic approach, on the other hand, allows us to capture the full spectral response of the microcavity for a range of different Rabi coupling strengths.

Finally, we use our exact results for the polariton wave function to obtain an estimate of polariton-polariton interactions that goes beyond previous calculations~\cite{Tassone1999,Rochat2000,MCombescot2007,MCombescot2008,Glazov2009}. In particular, we show that exchange interactions for a finite density of polaritons have a much smaller effect on the Rabi coupling than previously thought, due to the light-induced reduction of the exciton size in the very strong coupling regime.

\section{Model}

To describe a semiconductor quantum well \mmp{(or atomically thin material)} embedded in a planar optical cavity, we consider an effective two-dimensional model that includes light, matter, and the light-matter coupling:
\begin{align}
\hat H=\hat H_{\rm ph}+\hat H_{\rm mat}+\hat H_{\rm ph-mat}.
\label{eq:H}
\end{align}
The photonic part of the Hamiltonian is
\begin{align}
    \hat H_{\rm ph}=\sum_\k(\omega+\omega_{c\k}) c_\k^\dag c_\k.
\end{align}
Here, $c^\dagger_\k$ and $c_\k$ create and annihilate a cavity photon with in-plane momentum $\k$, while $\omega_{c\k}=k^2/2m_c$ is the 2D photon dispersion with $m_c$ the effective photon mass. For convenience, we write the cavity photon frequency at zero momentum, $\omega$, separately. Note that throughout this paper we work in units where $\hbar$ and the system area $A$ are both 1.

We consider the scenario where photons can excite electron-hole pairs across the band gap in the semiconductor, and these are in turn described by the effective low-energy Hamiltonian
\begin{align}
    \hat H_{\rm mat} = & \sum_\k\!\left[\left(\omega_{e\k}+\eg/2\right)\!e^\dag_\k
                 e_\k+\left(\omega_{h\k}+\eg/2\right)\!h^\dag_\k
                 h_\k\right]\nn \\
    &\hspace{-4mm}+\frac12\sum_{\k\k'\q}V(\q)\left[e^\dag_{\k+\q}e^\dag_{\k'-\q}e_{\k'}e_\k+h^\dag_{\k+\q}h^\dag_{\k'-\q}h_{\k'}h_\k \right. \nn \\ & \left. \hspace{19mm} -2e^\dag_{\k+\q}h^\dag_{\k'-\q}h_{\k'}e_\k\right],
\label{eq:Hmat}
\end{align}
where, for simplicity, we neglect spin degrees of freedom. $e^\dag_\k$ and $h^\dag_\k$ are the creation operators for electrons and holes at momentum $\k$, respectively, and the corresponding dispersions are $\omega_{e,h\k}=k^2/2m_{e,h}$ in terms of the effective masses $m_e$ and $m_h$. We explicitly include the electron-hole bandgap $\eg$ in the single-particle energies.

The interactions in a semiconductor quantum well are described by the (momentum-space) Coulomb potential $V(\q)=\frac{\pi}{m_r a_0 q}$, which we write in terms of the Bohr radius $a_0$ and the electron-hole reduced mass $m_r=(1/m_e+1/m_h)^{-1}$. In the absence of light-matter coupling, the Hamiltonian \eqref{eq:Hmat} leads to the existence of a hydrogenic series of electron-hole bound states, i.e., excitons, with energies
\begin{align}
\varepsilon_{n}=-\frac1{2m_ra_0^2}\frac1{(2n-1)^2}, \quad n=1,2,...,
\end{align}
which are independent of the pair angular momentum. Note that here and in the following we measure energies from the electron-hole continuum, or bandgap. Of particular interest are the circularly symmetric $s$ exciton states since these are the only ones that couple to light in our model. As a function of the electron-hole separation $r$, these exciton states have the wave functions~\cite{2DexcitonEnergy}
\begin{align}
    \Phi_{ns}(r)&=\frac{\sqrt{2}/a_0}{\sqrt{\pi(2n - 1)^3}}
  e^{-\frac{r/a_0}{2n - 1}} L_{n - 1}\left(\frac{2r/a_0}{2n - 1}\right),
\end{align}
%
where  $L_n$ are the Laguerre polynomials. Due to its importance in the following discussion, we will denote the binding energy of the $1s$ exciton as measured from the continuum by $\eb=1/(2m_ra_0^2)$, while we also note that its momentum-space wave function is
\begin{align}
    \tilde{\Phi}_{1s\k}=\frac{\sqrt{8\pi} a_0}{(1+k^2a_0^2)^{3/2}}.
\label{eq:phi1s}
\end{align}

Finally, the term $\hat H_{\rm ph-mat}$ describes the strong coupling of light to matter,
\begin{align}
    \hat H_{\rm ph-mat}  = & g\sum_{\k\q}\left[e^\dag_{\frac\q 2+\k} h^\dag_{\frac\q
                    2-\k}c_\q+c^\dag_\q h_{\frac\q 2-\k} e_{\frac\q 2+\k}\right]. \label{eq:Hg}
\end{align}
Here we have applied the rotating wave approximation, which should be valid when the light-matter coupling $g\ll a_0\eg$.  The form of Eq.~\eqref{eq:Hg} ensures that photons only couple to electron-hole states in $s$ orbitals, since the coupling strength $g$ is momentum independent\jfl{, an approximation which is similarly valid when $E_g$ greatly exceeds all other energy scales in the problem.}

\section{Renormalization of the cavity photon frequency}
\label{sec:renorm}

\begin{figure}[t]
\centering
\includegraphics[width=\columnwidth]{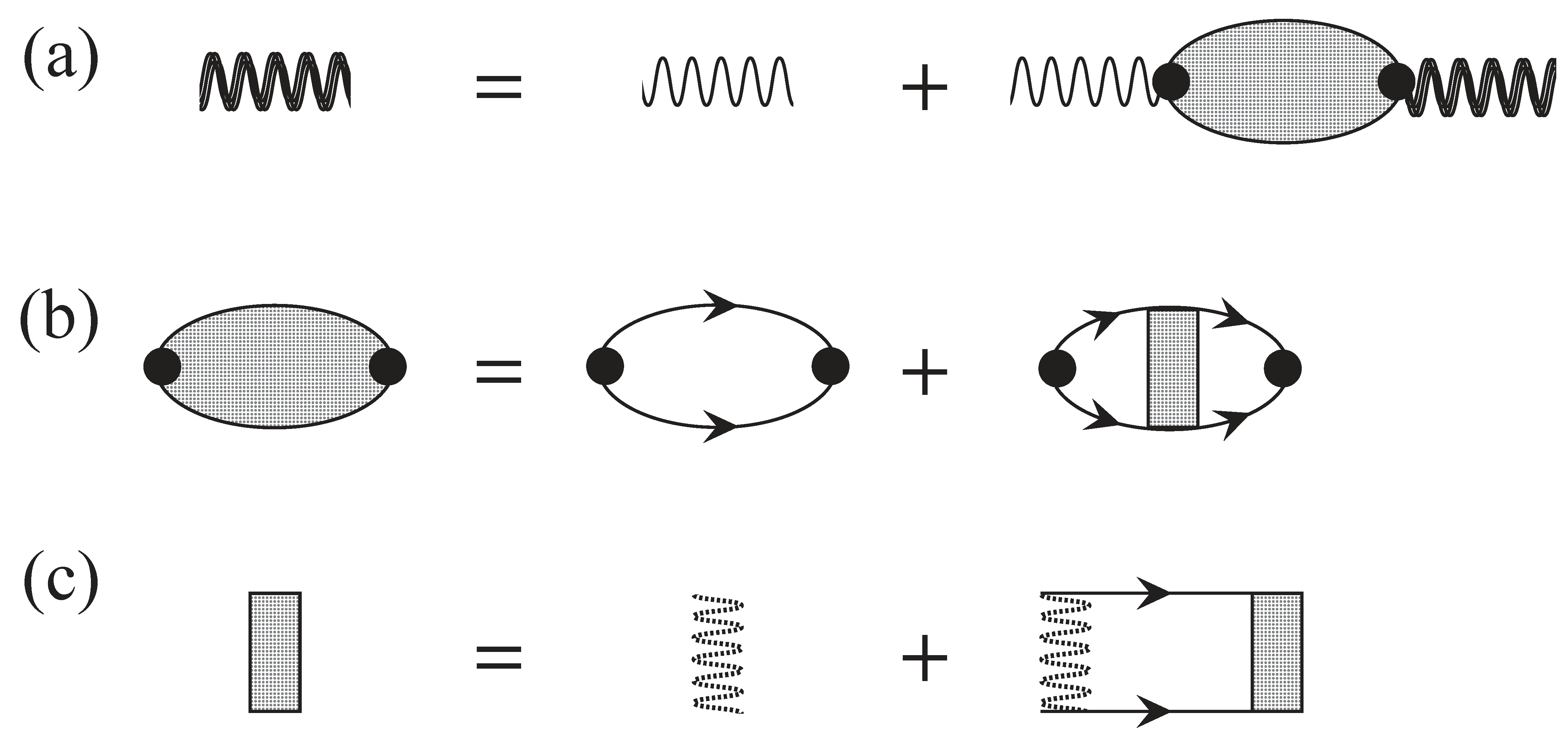}
\caption{Feynman diagrams involved in the renormalization of the cavity photon frequency. (a) Dyson equation for the photon propagator (double wavy line) in terms of the bare photon propagator (single wavy line) and the photon self energy (shaded ellipse). The black circles indicate the light-matter coupling $g$. (b) Separation of the self energy into the bare electron-hole polarization bubble (left) and a term containing any positive number of repeated Coulomb interactions (right). The lines represent bare electrons and holes, and the square represents the electron-hole $T$ matrix. (c) The equation satisfied by the $T$ matrix, where the dotted wavy line represents the Coulomb interaction.}
\label{fig:diagrams}
\end{figure}

We now show how the light-matter coupling in the microscopic model leads to an arbitrarily large shift of the bare cavity photon frequency $\omega$, thus necessitating a renormalization procedure akin to that used in quantum electrodynamics (see Fig.~\ref{fig:diagrams}).  Since the argument is independent of the photon momentum, we focus on photons at normal incidence, i.e., at zero momentum in the plane, and we relegate the details of the finite-momentum case to Appendix~\ref{app:FiniteQ}.

To illustrate the need for renormalization in the simplest manner possible, we start with a single photon in the absence of light-matter coupling, corresponding to the state $c_\0^\dag\ket{0}$ with cavity frequency $\omega$, where $\ket{0}$ represents the vacuum state for light and matter. We then use second-order perturbation theory to determine the shift in the cavity photon frequency for small coupling $g$,
\begin{align} \label{eq:PTshift}
    \Delta \omega \simeq g^2 \sum_{n=1}^\infty \frac{|\Phi_{ns}(0)|^2}{\omega - \varepsilon_{n}-\eg} + g^2\sum_\k \frac{1}{\omega -\ok - \eg} ,
\end{align}
where $\ok\equiv \omega_{e\k} + \omega_{h\k} = k^2/2m_r$.  The first term on the right hand side of \eqref{eq:PTshift} is a convergent sum involving the exciton bound states in the $s$-wave channel, while the second term results from unbound electron-hole pairs, i.e., $e^\dag_\k h^\dag_{-\k}\ket{0}$ for a given relative momentum $\k$, where we have neglected scattering induced by the Coulomb interaction \jfl{since our arguments in the following hinge on the high-energy behavior where this is negligible}.  We can immediately see that the momentum sum diverges, unless we impose a UV momentum cutoff $\Lambda$, in which case we have $\Delta\omega \sim -g^2 m_r \ln\Lambda$.  Note that $g$ is a well-defined coupling constant since the corrections to the exciton energies do not depend on $\Lambda$. Indeed, for the ground-state exciton, the lowest-order energy shift due to light-matter coupling is
\begin{align} \label{eq:Xshift}
    \Delta\varepsilon_1 = \frac{g^2 |\Phi_{1s}(0)|^2}{\eg + \varepsilon_1 - \omega} ,
\end{align}
which resembles what one would expect from the simple exciton-photon model~\cite{Hopfield1958,Microcavities}.

Physically, the cutoff $\Lambda$ is determined by a high-energy scale of the system such as the crystal lattice spacing, which is beyond the range of validity of our low-energy microscopic Hamiltonian. Therefore, we require a renormalized cavity photon frequency that is independent of the high-energy physics associated with $\Lambda$. Indeed, this is reminiscent of the UV divergence occurring in the vacuum polarization of quantum electrodynamics \mmp{(Fig.~\ref{fig:diagrams})}, which leads to a screening of the electromagnetic field~\cite{WeinbergVol1}. We emphasize that the emergence of a UV divergence should not be interpreted as a failure of perturbation theory \cite{Boyd1999}, but persists in the full microscopic theory as we now demonstrate.

To proceed, we consider the most general wave function for a single polariton at zero momentum
\begin{align}
    \ket{\Psi} &=\sum_\k \varphi_\k e^\dag_\k h^\dag_{-\k}\ket{0}+\gamma c_\0^\dag \ket{0}.
    \label{eq:psi}
\end{align}
Here, and in the following, we assume that the state is normalized, i.e., $\braket{\Psi}{\Psi}=\sum_\k |\varphi_\k|^2+|\gamma|^2 =1$.  We then project the Schr\"{o}dinger equation, $(E +\eg-\hat H)\ket{\Psi} = 0$, onto photon and electron-hole states, which yields the coupled equations
\begin{subequations}\label{eq:vareqs}
\begin{align}
        (E-\ok)\varphi_\k &=-\sum_{\k'}V(\k-\k')\varphi_{\k'} +g\gamma, \label{eq:i}\\ 
        (E-\omega+\eg)\gamma & = g\sum_\k \varphi_\k,
        \label{eq:ii}
\end{align}
\end{subequations}
where we have again defined the energy $E$ with respect to the bandgap energy $\eg$.

Defining $\beta_\k \equiv \sum_{\k'}V(\k-\k')\varphi_{\k'}/(-E+\ok)$ and rearranging Eq.~\eqref{eq:i} gives the electron-hole wave function 
\begin{align}\label{eq:phi}
\varphi_\k = \beta_\k +\frac{g\gamma}{E-\ok} .
\end{align}
In the absence of light-matter coupling, the lowest-energy excitonic solution corresponds to the $1s$ exciton state, i.e., $\varphi_\k = \beta_\k = \tilde{\Phi}_{1s\k}$. However, once $g\neq 0$, we see that the electron-hole wave function is modified by the photon, even in the limit of small $g$. Light-induced changes to the exciton radius have previously been predicted within approximate variational approaches~\cite{Khurgin2001,Zhang2013,Yang2015} and have already been observed in experiment~\cite{Brodbeck2017}.  Our exact treatment shows that even the functional form of the exciton wave function changes, since the second term in Eq.~\eqref{eq:phi} yields $\varphi(r)\sim g\gamma m_r\ln(r)/\pi$ in real space as $r \to 0$. By contrast, the function $\beta(r)$ is regular at the origin (see Appendix~\ref{app:2DGreenfunction}), and hence the definition Eq.~\eqref{eq:phi} serves to isolate the divergent short-range behavior of the electron-hole wave function.

The short-distance behavior of the real-space exciton wave function is intimately connected to the renormalization of the bare cavity photon frequency. To see this, we rewrite Eq.~\eqref{eq:vareqs} in terms of $\beta_\k$:
\begin{subequations}\label{eq:betaeqs}
\begin{align}
    &(E-\ok)\beta_\k  =-\sum_{\k'}V(\k-\k')\beta_{\k'}+g\gamma \sum_{\k'}\frac{V(\k-\k')}{-E+\bar\omega_{\k'}},
    \label{eq:ib}\\
    &\left(E-\omega+\eg+g^2\sum_\k\frac1{-E+\ok}\right) \gamma  = g\sum_\k \beta_\k.
    \label{eq:iib}
\end{align}
\end{subequations}
Here, one can show that all the sums are convergent except for the sum on the left-hand side of Eq.~\eqref{eq:iib}, which displays the same logarithmic dependence on the UV cutoff $\Lambda$ as in Eq.~\eqref{eq:PTshift}.  Thus, we have isolated the high-energy dependence, which can now be formally removed by relating the bare parameter $\omega$ to the physical cavity photon frequency observed in experiment.

\mmp{We emphasize that the precise renormalization procedure depends on the specific low-energy model under consideration. In particular, if we approximate the electron-hole interactions as heavily screened and short-ranged, then one can show that the lowest-order shift in the exciton energy in Eq.~\eqref{eq:Xshift} also contains a UV divergence. Hence, in this case one finds that the light-matter coupling $g$ must vanish logarithmically with $\Lambda$, while the cavity frequency retains its bare value.}

\subsection{Relation to experimental observables}
\label{sec:relation_to_two_level}
Experimental spectra are typically fitted using phenomenological two-level exciton-photon models in order to extract the polariton parameters.  Therefore, we must recover such a two-level model from Eq.~\eqref{eq:betaeqs} in order to relate the bare parameters in our microscopic description to the observables in experiment.  This is easiest to do in the regime $g\ll a_0\eb$, which is what exciton-photon models already implicitly assume~\cite{Hopfield1958}. In this limit, we can assume that the convergent part of the exciton wave function is unchanged such that $\beta_\k\simeq\beta \tilde{\Phi}_{1s\k}$, where $\beta$ is a complex number.

Applying the operator $\sum_\k \tilde{\Phi}_{1s\k}\{\cdot\}$ to Eq.~\eqref{eq:ib} and using the \sch equation for $\tilde{\Phi}_{1s\k}$, we obtain
\begin{align} \label{eq:2lvla}
(E+\eb)\beta & \simeq g\gamma
               \sum_{\k,\k'}\frac{V(\k-\k')\tilde{\Phi}_{1s\k}}{\eb+\bar{\omega}_{\k'}} = g\gamma\Phi_{1s}(0),
\end{align}
where we have taken $E\simeq -\eb$ in the intermediate step since the energies of interest are close to the exciton energy in this limit. Similarly, we can approximate Eq.~\eqref{eq:iib} as
\begin{align}
    \left(E-\omega+\eg+g^2\sum_\k\frac1{\eb+\ok}\right) \gamma \simeq g\beta \Phi_{1s}(0) .
\end{align}
If we now identify the Rabi coupling as 
\begin{align}\label{eq:Omega_def}
\Omega &\equiv g\Phi_{1s}(0)=\frac{g}{a_0}\sqrt{\frac2\pi},
\end{align}
and the (finite) physical photon-exciton detuning as
\begin{align}\label{eq:detuning_renor}
\delta\equiv\omega-E_g-g^2\sum_\k\frac1{\eb+\ok}+\eb,
\end{align}
we arrive at the following simple two-level approximation of Eq.~\eqref{eq:betaeqs}:
\begin{align}\label{eq:2lvl}
    E 
    \begin{pmatrix}
    \beta \\ 
    \gamma
    \end{pmatrix}
    = 
    \begin{pmatrix}
    -\eb & \Omega \\ 
    \Omega & \delta - \eb
    \end{pmatrix}
    \begin{pmatrix}
    \beta \\ 
    \gamma
    \end{pmatrix} .
\end{align}
This yields the standard solutions for the lower (LP) and upper (UP) polaritons, with corresponding energies:
\begin{align}\label{eq:two-level dispersion}
E_{{\tiny  \begin{matrix} \text{LP}\\\text{UP}\end{matrix}}} &=-\eb+\frac12\left(\delta\mp\sqrt{\delta^2+4\Omega^2}\right),
\end{align}
and photon Hopfield coefficients 
\begin{align} \label{eq:hop}
    \gamma_{{\tiny  \begin{matrix} \text{LP}\\\text{UP}\end{matrix}}} 
    = \mp \sqrt{\frac{1}{2} \left(1\mp \frac{\delta}{\sqrt{\delta^2+4\Omega^2}} \right)} .
\end{align}
Note that the (positive) exciton Hopfield coefficent is simply $\beta = \sqrt{1-|\gamma|^2}$, and $|\gamma|^2$ is the photon fraction.

With the identifications \eqref{eq:Omega_def} and \eqref{eq:detuning_renor}, the set of equations~\eqref{eq:betaeqs} (or equivalently \eqref{eq:vareqs}) now represents a fully renormalized problem where the momentum cutoff can be taken to infinity without affecting the low-energy properties.  While the exciton-photon Rabi coupling \eqref{eq:Omega_def} has been defined previously~\cite{Tassone1999}, the detuning \eqref{eq:detuning_renor} differs from previous work~\cite{Byrnes2010,Kamide2010,Kamide2011,Yamaguchi2013,Xue2016}, which only considered the bare cavity photon frequency $\omega$.

\subsection{Diagrammatic approach to renormalization}
\label{sub:diag}
To provide further insight into the origin of the photon renormalization, we now present an alternative derivation in terms of Feynman diagrams, as illustrated in Fig.~\ref{fig:diagrams}.  The key point is that the energy spectrum produced by Eq.~\eqref{eq:betaeqs} (or Eq.~\eqref{eq:vareqs}) may also be determined from the poles of the (retarded) photon propagator once this is appropriately dressed by light-matter interactions.

We first note that the photon propagator $G_{\rm C}$ satisfies the Dyson equation~\cite{fetterbook} in Fig.~\ref{fig:diagrams}(a):
\begin{align}
    G_{\rm C}(E) & = G_{\rm C}^{(0)}(E)+G_{\rm C}^{(0)}(E)\Sigma(E)G_{\rm C}(E) \nn \\ & =\frac1{[G_{\rm C}^{(0)}(E)]^{-1}-\Sigma(E)},
    \label{eq:GC}
\end{align}
where
\begin{align}
    G_{\rm C}^{(0)}(E) & = \frac1{E-\omega+\eg}
\end{align}
is the photon propagator in the absence of light-matter coupling (we remind the reader that we measure energy from $\eg$), and $\Sigma$ is the self energy. Throughout this subsection, we will assume that the energy contains a positive imaginary infinitesimal that shifts the poles of the photon propagator slightly into the lower half plane.  For simplicity, we again consider the photon at normal incidence --- our arguments are straightforward to generalize to finite momentum.

\begin{figure*}
\centering
\includegraphics[width=2\columnwidth]{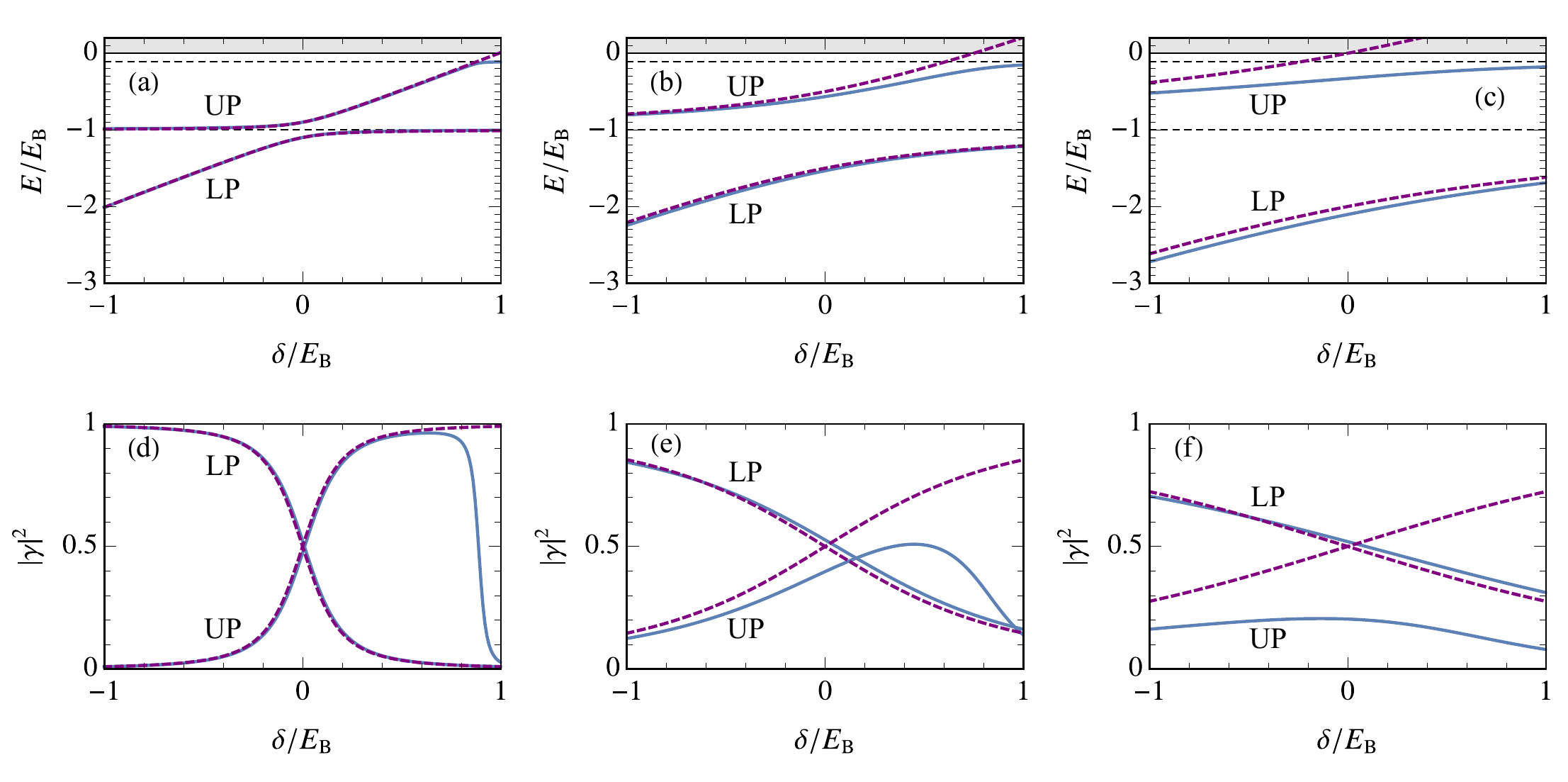}
\caption{Energies (top) and photon fractions (bottom) of the lower and upper polaritons obtained within our microscopic model in Eq.~\eqref{eq:vareqs} (blue solid lines) and within the two-level model \eqref{eq:2lvl} (purple dashed lines). We show the results as a function of detuning, and from left to right we consider increasing strength of the Rabi coupling: (a,d) $\Omega/\eb=0.1$; (b,e) $\Omega/\eb=0.5$; (c,f) $\Omega/\eb=1$. In (a-c) the horizontal dashed lines are the 1$s$ and 2$s$ exciton states with energies $\varepsilon_1=-\eb$ and $\varepsilon_2=-\eb/9$, respectively, while the shaded regions correspond to the electron-hole continuum.} 
\label{fig:energies}
\end{figure*}

As shown in Fig.~\ref{fig:diagrams}(b,c), the photon self energy $\Sigma$ arises from all possible scattering processes involving the excitation of an electron-hole pair.  This can be written as the sum of two terms
\begin{align}
    \Sigma(E) & = \Sigma^{(1)}(E) + \Sigma^{(2)}(E) ,
    \label{eq:photonselfenergy}
\end{align}
with
\begin{subequations}
\label{eq:Sigma}
\begin{align}
    \Sigma^{(1)}(E)&=
        g^2\sum_{\k}\frac1{E-\bar\omega_\k},\\
    \Sigma^{(2)}(E)&=g^2\sum_{\k,\k'}\frac{T(\k,\k';E)}{(E-\bar\omega_\k)(E-\bar\omega_{\k'})}.
\end{align}
\end{subequations}
Here, we see that $\Sigma^{(1)}$ is cutoff dependent, while $\Sigma^{(2)}$ is well behaved and depends on the electron-hole $T$ matrix $T(\k,\k';E)$ at incoming and outgoing relative momenta $\k$ and $\k'$, respectively. Note that the $T$ matrix only depends on the Coulomb interaction and can be completely determined in the absence of light-matter coupling.

We are now in a position to find the spectrum of the dressed photon propagator. From its definition, Eq.~\eqref{eq:GC}, we see that the poles satisfy
\begin{align}
    & E-\omega+\eg+\sum_{\k}\frac{g^2}{-E+\bar\omega_\k}= \sum_{\k,\k'}\frac{g^2T(\k,\k';E)}{(E-\bar\omega_\k)(E-\bar\omega_{\k'})}.
    \label{eq:photonpole}
\end{align}
This expression is very reminiscent of Eq.~\eqref{eq:iib}, and indeed we demonstrate in Appendix \ref{app:2DGreenfunction} that Eq.~\eqref{eq:betaeqs} directly leads to Eq.~\eqref{eq:photonpole}. While the right hand side of Eq.~\eqref{eq:photonpole} is convergent (see Appendix~\ref{app:2DGreenfunction}), the sum on the left hand side again depends logarithmically on the cutoff $\Lambda$.  This necessitates the redefinition of the cavity photon frequency to cancel this high-energy dependence, in the same manner as in Eq.~\eqref{eq:detuning_renor}.

While all properties of the electron-hole-photon system can equally well be extracted from the diagrammatic approach, we find it more transparent in the following to work instead with wave functions. For completeness, we mention that the integral representation of the electron-hole Green's function was derived in Ref.~\cite{2DGreenfunction}, and we use this in Appendix~\ref{app:2DGreenfunction} to obtain an integral representation of the $T$ matrix.

\section{Exact results for a single exciton-polariton}
\label{sec:comparison_numerics}
We now turn to the polariton eigenstates and energies within our low-energy microscopic Hamiltonian \eqref{eq:H}.  We obtain exact results by numerically solving the linear set of equations \eqref{eq:vareqs} using the substitutions \eqref{eq:Omega_def} and \eqref{eq:detuning_renor} to relate the bare microscopic parameters to the physical observables in the two-level model \eqref{eq:2lvl}.  Even though the unscreened Coulomb interaction features a pole at $q=0$, we emphasize that this is integrable, and thus Eq.~\eqref{eq:vareqs} can be solved using standard techniques for Fredholm integral equations~\cite{numericalrecipes}.

Figure~\ref{fig:energies} shows how our numerical results compare with the energies and photon fractions of the upper and lower polariton states predicted by the two-level model \eqref{eq:2lvl}.  In the regime $\Omega \ll \eb$, we find good agreement between the two models, where the LP and UP states respectively correspond to the first and second lowest energy eigenstates of Eq.~\eqref{eq:vareqs}.  Note that there will be deviations when the UP state approaches the electron-hole continuum at large photon-exciton detuning, since the two-level model cannot capture the fact that the UP evolves into the $2s$ exciton state in this limit --- see the energy in the limit of large detuning in Fig.~\ref{fig:energies}(a), and the associated strong suppression of the photonic component in Fig.~\ref{fig:energies}(d).

\begin{figure*}[t] \centering \includegraphics[width=\textwidth]{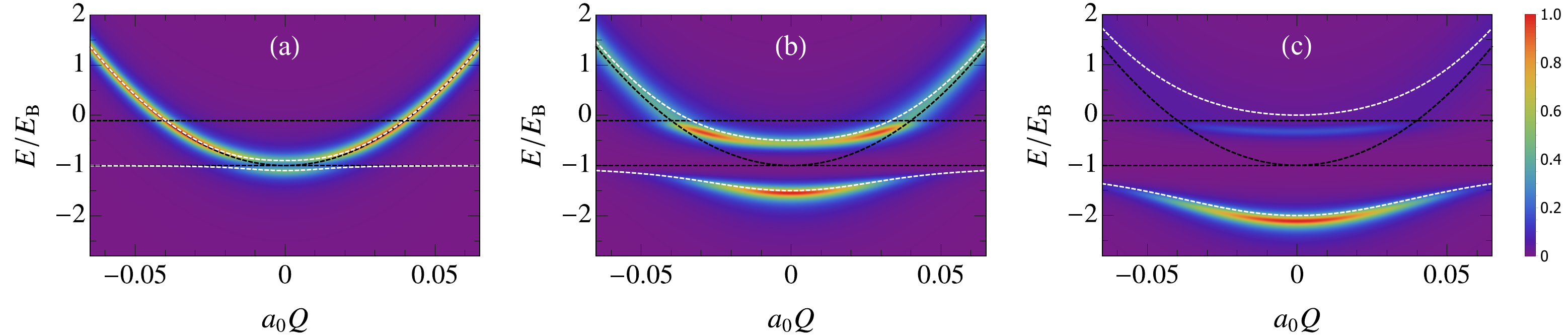} \caption{ Transmission spectrum $\mathcal{T}(\Q,E)$ of a single photon at detuning $\delta=0$ and for Rabi couplings (a) $\Omega/\eb=0.1$, (b) $\Omega/\eb=0.5$, and (c) $\Omega/\eb=1$.  The polariton dispersions predicted from the two-level model in Eq.~\eqref{eq_energy_mat} are shown as white dashed lines, while the bare photon and $1s$ and $2s$ exciton dispersions are shown as dashed black lines.  The masses are taken to be the realistic values \cite{EffectiveMass} $m_e=0.063\,m_{0}, m_h=0.5\,m_{0}, m_c=10^{-4}\, m_{0}$, with $m_{0}$ the free electron mass.  To enhance visibility over the large energy range displayed in the figures, we are taking the photon lifetime broadening to be $\Gamma=0.1\,\eb$, which exceeds typical values in high-quality-factor microcavities. Each spectrum has been normalized to the maximum transmission.}
\label{fig:transmission}
\end{figure*}

\mmp{The regime of very strong light-matter coupling, where $\Omega \sim \eb$, can now be achieved in a range of materials~\cite{Sanvitto2016,Guillet2016}.  In this case,} the two-level model fails to describe the UP state even close to resonance, $\delta = 0$, since the photon becomes strongly coupled to higher energy exciton states and the electron-hole continuum.  This manifests itself as a suppression of the photon fraction in the second lowest eigenstate of Eq.~\eqref{eq:vareqs}, as displayed in Fig.~\ref{fig:energies}(e,f). Such behavior cannot be captured by simply including the 2$s$ exciton state, like in the case of weaker Rabi coupling. Indeed, we find that it cannot even be fully described with a model that includes all the exciton bound states, thus highlighting the importance of the electron-hole continuum.  However, note that for the parameters shown in Fig.~\ref{fig:energies}(a-c), the UP state obtained within the microscopic model always lies below the continuum, and it thus still corresponds to a discrete line in the energy spectrum (if we neglect photon loss and sample disorder).

To further elucidate the behavior of the polariton states, we calculate the spectral response that one can observe in experiment. Specifically, we focus on the scenario where the detuning is tuned to resonance (i.e., $\delta = 0$) and the polariton momentum $\Q$ is varied. Thus, we solve the equations for the finite center-of-mass case in Appendix~\ref{app:FiniteQ} and then numerically obtain the photon propagator at momentum $\Q$ and energy $E$:
\begin{align}\label{eq:photon_green}
  G_{\rm C}(\Q,E) =  \sum_n \frac{{|\gamma_n^{(\Q)}|}^2}{E-E^{(\Q)}_n+i \Gamma}.
\end{align}
Here $\Gamma$ accounts for the finite lifetime effects in the microcavity, while $|\gamma_n|^2$ and $E_n$ correspond to the photon fraction and energy, respectively, of the $n$th eigenstate in the finite-momentum problem, Eq.~\eqref{eq:finiteQ}. The different spectral responses can then be directly extracted from $G_{\rm C}(\Q,E)$~\cite{CiutiPRA2006,Cwik2016}.  The spectral function has also been previously calculated in Ref.~\cite{Citrin2003}, but within an unrenormalized theory.

In Fig.~\ref{fig:transmission} we display the photon transmission spectrum 
$\mathcal{T}(\Q,E)\propto {|G_{\rm C}(\Q,E)|}^2$ --- note that the precise proportionality factor depends on the details of the microcavity~\cite{CiutiPRA2006,Cwik2016}.  When $\Omega \ll \eb$, we see that we recover the usual polariton dispersions predicted by the two-level model, even for energies $E>0$ where there is no bound exciton.  However, for larger Rabi couplings, the UP line at positive energies becomes progressively broadened by interactions with the continuum of unbound electron-hole states, while for negative energies it becomes strongly modified by the $2s$ exciton state, as shown in Fig.~\ref{fig:transmission}(b,c).  In particular, we observe that the spectral weight (or oscillator strength) of the UP state is significantly reduced, as has been observed in experiment.  This reduction in UP oscillator strength is due to the transfer of the photonic component to other light-matter states involving both the electron-hole continuum as well as the higher energy exciton bound states.  Indeed, in the very strong coupling regime, one can in principle probe Rydberg exciton-polaritons involving the exciton states of larger $n$~\cite{Kazimierczuk2014}.  Note that the behavior in Fig.~\ref{fig:transmission} implicitly depends on the electron-hole reduced mass as well as the exciton mass, which is beyond the two-level model.

\begin{figure*}[t] \centering
  \includegraphics[width=\textwidth]{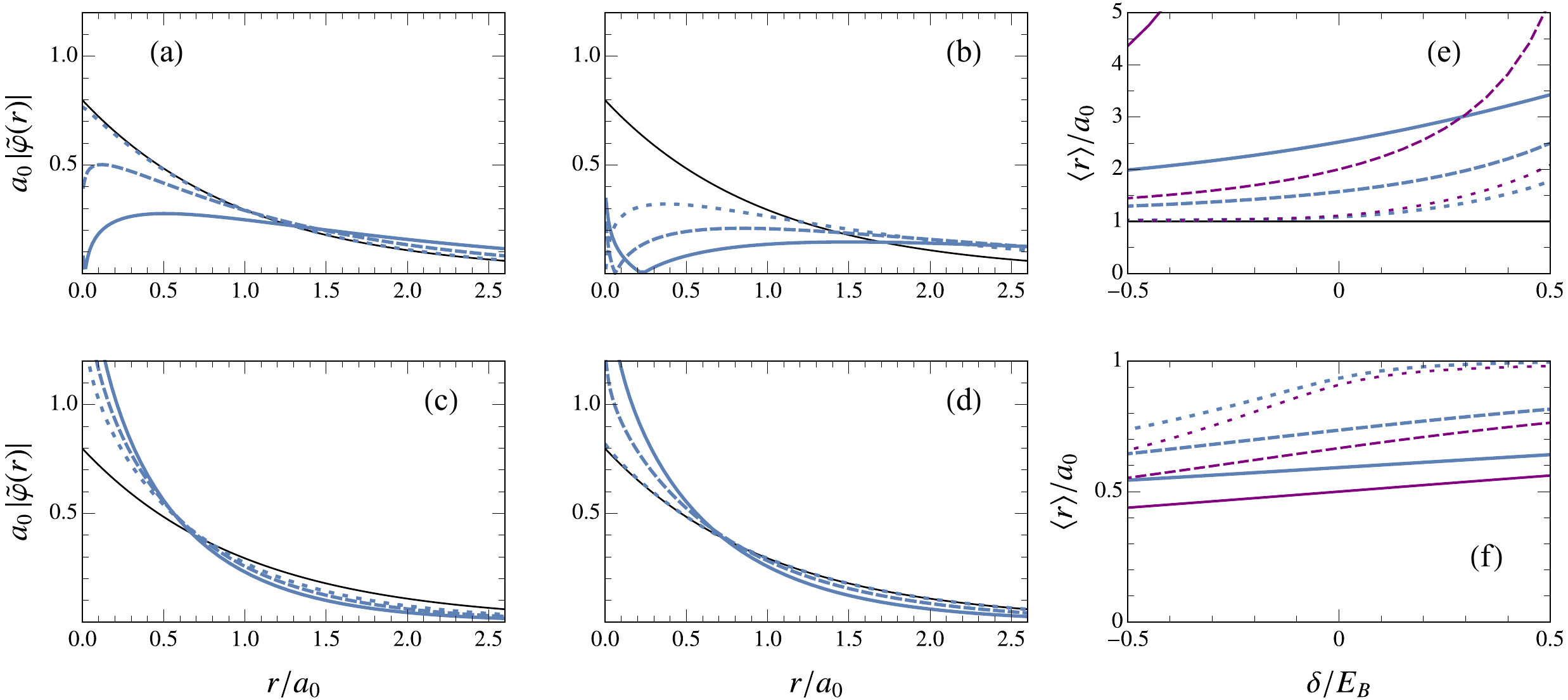} 
\caption{(a-d) Normalized electron-hole wave function as a function of separation, and (e,f) the mean electron-hole separation as a function of detuning.  The upper (lower) panels correspond to the upper (lower) polariton, and we take $\delta/\eb=-0.5$ in (a,c) and $\delta/\eb=0.5$ in (b,d).  We show the exact results obtained from our microscopic theory (blue lines) for the Rabi coupling strengths of $\Omega/\eb=0.1$ (dotted line), $\Omega/\eb=0.5$ (dashed line), and $\Omega/\eb=1$ (solid line).  Note that the numerically calculated wave functions in (a-d) are only shown for $r>0.005\,a_0$.  We also show the $1s$ exciton wave function and average separation (thin black line).  In (e,f) the purple lines correspond to the result \eqref{eq:variational} from the variational approach~\cite{Khurgin2001,Zhang2013}, where we take the detuning to be $\tilde\delta$ (see main text).}
\label{fig:wave}
\end{figure*}

In contrast to the upper polariton, the spectral properties of the lower polariton are remarkably close to those predicted by the two-level model in the regime $\Omega \sim \eb$, as shown in Figs.~\ref{fig:energies} and \ref{fig:transmission}.  This is not unreasonable since the LP state is well separated in energy from all the excited electon-hole states. On the other hand, Eq.~\eqref{eq:phi} implies that the electron-hole wave function of the polariton becomes strongly modified by the coupling to light. To confirm this, we plot in Fig.~\ref{fig:wave}(a-d) the (normalized) real-space electron-hole wave function at zero center of mass momentum, corresponding to $\tilde\varphi(r)=\tfrac1{\sqrt{1-|\gamma|^2}}\sum_\k e^{i\k\cdot\r}\varphi_\k$.  For both LP and UP states, we see that $\tilde\varphi(r)$ only resembles the ground-state exciton wave function when $\Omega \ll \eb$ and the photon fraction is small, $|\gamma| \ll 1$. However, we always have $\tilde\varphi(r) \to \frac{g\gamma m_r}{\pi\sqrt{1-|\gamma|^2}}\ln(r)$ when $r \to 0$, as expected from Eq.~\eqref{eq:phi}.  Moreover, the coupling to light can significantly shrink (expand) the size of the bound electron-hole pair in the lower (upper) polariton, as shown in Fig.~\ref{fig:wave}(e,f).

Light-induced modifications to the exciton size have also been investigated using a variational approach~\cite{Khurgin2001}, where the electron-hole wave function is assumed to have the same form as the 1$s$ exciton state, but with a different Bohr radius $a_0/\lambda$. In general, such a wave function is a superposition of an infinite number of $s$-orbital exciton and continuum states, and is thus closer to our approach than the simple two-level model. Minimizing the energy with respect to the variational parameter $\lambda$ yields the usual two-level Hamiltonian \eqref{eq:2lvl} but with a shifted photon-exciton detuning $\tilde{\delta} = \delta -\Omega^2/\eb$~\cite{Zhang2013}.  Therefore, once we identify $\tilde{\delta}$ as the physical detuning, the variational approach recovers the standard polariton states, even though the exciton radius is changed, such that
\begin{align}\label{eq:variational}
    \lambda = 1 - \frac{\gamma}{\sqrt{1-|\gamma|^2}} \frac{\Omega}{\eb} ,
\end{align}
where $\gamma$ is given by Eq.~\eqref{eq:hop} with $\delta$ replaced by the new detuning $\tilde{\delta}$.  Thus we see that the light-induced changes to the exciton wave function shift the cavity photon frequency in a manner that resembles the renormalization appearing in the exact problem, although the shift is independent of the UV momentum cutoff since the variational wave function is chosen to be regular at the origin.  This formal similarity between the exact and variational approaches provides a possible explanation for why the exact LP state is so well approximated by the two-level model in Figs.~\ref{fig:energies} and \ref{fig:transmission} despite having a strongly modified electron-hole wave function.

\begin{figure*}[ht]
\centering
\includegraphics[width=2\columnwidth]{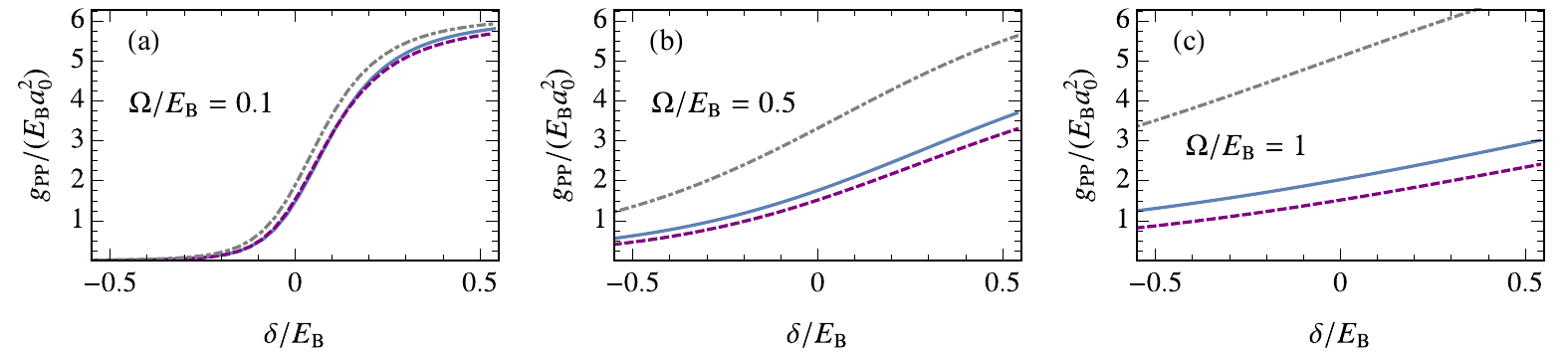} 
\caption{Polariton-polariton interaction strength $g_{\rm PP}$ for several values of the light-matter coupling: (a) $\Omega/E_\mathrm{B}=0.1$;  (b) $\Omega/E_\mathrm{B}=0.5$; and (c) $\Omega/E_\mathrm{B}=1$. We compare our numerical results obtained from Eq.~\eqref{eq:gpp2} (blue solid lines) with the simple exciton approximation $g^{(0)}_{\rm PP}=\beta^4 g_{\rm XX}$ (purple dashed lines) \mmp{and the commonly accepted interaction strength that includes the exciton oscillator strength saturation correction \eqref{eq:corra}: $g_{\rm PP}^{\rm sat}=g^{(0)}_{\rm PP}+\frac{32\pi}{7}\Omega a_0^2\beta^3|\gamma|$ (gray dash-dotted line).}}
  \label{fig:gpp}
\end{figure*}

In Fig.~\ref{fig:wave}(e,f), we see that the variational approach \eqref{eq:variational} qualitatively captures how the exciton radius changes with Rabi coupling and detuning. However, it gives $\lambda = 0$, i.e., a diverging radius, for the upper polariton when the UP energy in Eq.~\eqref{eq:two-level dispersion} crosses zero, and it thus fails to describe the UP as it approaches the electron-hole continuum. Moreover, even in the case of the LP, the variational approach noticeably overestimates the change in the exciton size in the very strong coupling regime.

Finally we note that, unlike Ref.~\cite{Kamide2010}, we do not obtain strongly bound electron-hole pairs with a radius that is far smaller than the Bohr radius $a_0$ for experimentally relevant parameters.  Consequently, we conclude that the electron-hole pairs in the presence of photons should not be considered as Frenkel excitons. We attribute this major qualitative difference to the unrenormalized nature of the theory considered in Ref.~\cite{Kamide2010}.

\section{Polariton-polariton interaction}
We now demonstrate how our microscopic model of exciton-polaritons allows us to go beyond previous calculations~\cite{Tassone1999,Rochat2000,MCombescot2007,MCombescot2008,Glazov2009} of the interaction strength $g_{\rm{PP}}$ for spin-polarized lower polaritons at zero momentum. An accurate knowledge of this interaction strength is fundamental to understanding the many-body polariton problem, both in the mean-field limit~\cite{Deng2010,Carusotto2013} and beyond.  However, its precise value has been fraught with uncertainty, as experiments have reported values differing by several orders of magnitude \cite{FerrierPRL2011,Brichkin2011,Rodriguez2016,WalkerPRL2017, Estrecho2018}.  Moreover, there are conflicting theoretical results regarding the effect of the light-matter coupling~\cite{Tassone1999,Rochat2000,MCombescot2007,Xue2016}.

The key advantage of our approach is that we know the full polariton wave function including the modification of the relative electron-hole motion, as illustrated in Fig.~\ref{fig:wave}. This allows us to obtain non-trivial light-induced corrections to the polariton-polariton interaction strength. To proceed, we introduce the exact
zero-momentum lower polariton operator,
\begin{align}
P^\dag = \sum_\k \varphi_{\rm{LP}\k} e^\dag_\k h^\dag_{-\k} + \gamma_{\rm{LP}} a^\dag,
\label{eq:pol}
\end{align}
where the subscripts on $\varphi_{\rm{LP}\k}$ and $\gamma_{\rm{LP}}$ indicate that these form the (normalized) solution of Eq.~\eqref{eq:vareqs} at $E=E_{\rm{LP}}$, where $E_{\rm LP}$ is the exact ground-state energy for one polariton.  As in previous work~\cite{Tassone1999,Rochat2000,MCombescot2007,MCombescot2008,Glazov2009}, we restrict ourselves to the Born approximation of polariton-polariton interactions. In this case, the total energy of the two-polariton system is
\begin{align}
    \varepsilon_{\rm tot} = \frac{\bra{PP}\hat H \ket{PP}}{\langle PP | PP \rangle} = 2E_g + 2 E_{\rm LP} + \frac{g_{\rm PP}}A ,
\end{align}
where $\ket{PP}\equiv P^\dag P^\dag\ket{0}$ and we have reinstated the system area $A$.  The second term corresponds to the interaction energy $g_{\rm PP} N(N-1)/2A$ for $N=2$ identical bosons. Rearranging then gives
\begin{align} \label{eq:gpp}
    \frac{g_{\rm PP}}A = \frac{1}2 
    \bra{PP} \hat H-2E_g-2E_{\rm{LP}}\ket{PP}.
\end{align}
The form of Eq.~\eqref{eq:gpp} accounts for the fact that the polariton operator only approximately satisfies bosonic commutation relations due to its composite nature~\cite{Tassone1999,MCombescot2008}. Specifically, we note that $\braket{PP}{PP} = 2 + O(a_0^2/A)$ and then only keep terms up to $O(a_0^2/A)$.

Evaluating Eq.~\eqref{eq:gpp} and using Eq.~\eqref{eq:vareqs} to eliminate $\gamma_{\rm{LP}}$ (for details see Appendix~\ref{app:gPP}) we arrive at
\begin{align}
    g_{\rm{PP}} = 2 \sum_\k \left(\ok \! -\! E_{\rm{LP}} \right) \varphi_{\rm{LP}\k}^4 \!-\! 2 \sum_{\k,\k'} V(\k\!-\!\k') \varphi_{\rm{LP}\k}^2 \varphi_{\rm{LP}\k'}^2.
\label{eq:gpp2}
\end{align}
This result is rather appealing, since it has the exact same functional form as in the exciton limit~\cite{Tassone1999} (see also Appendix~\ref{app:gPP}), which is simply obtained by taking $\varphi_{\rm{LP}\k}\to \tilde\Phi_{1s\k}$ and $E_{\rm{LP}}\to -\eb$. Indeed, in this limit we recover the accepted value for the exciton-exciton interaction strength: $g_{\rm{PP}}\to g_{\rm XX}\simeq3.03/m_r=6.06\eb a_0^2$~\cite{Tassone1999}.

The leading order effect of light-matter coupling is the hybridization of excitons and photons, resulting in $g^{(0)}_{\rm PP}=\beta^4\,g_{\rm XX}$, where $\beta$ is the Hopfield exciton coefficient in Eq.~\eqref{eq:2lvl}.  Beyond this, one expects corrections involving powers of $\Omega/\eb$ (or equivalently $\Omega a^2_0$) in the very strong coupling regime. Previous theoretical work implicitly focused on the lowest order correction, referred to as the ``oscillator strength saturation''~\cite{Brichkin2011}, which corresponds to a positive addition to $g_{\rm{PP}}^{(0)}$ that is proportional to $\Omega a_0^2$.  However, there are conflicting predictions for the proportionality factor~\cite{Tassone1999,Rochat2000,MCombescot2007}, and all these predictions are derived under the assumption that the relative electron-hole wave function remains unchanged in the presence of light-matter coupling.  By contrast, our approach provides a complete determination of $g_{\rm{PP}}$ within the Born approximation, and hence all the previously obtained corrections are captured in Eq.~\eqref{eq:gpp2}, along with additional terms that contribute at the same order and higher than $\Omega a_0^2$ (see Appendix~\ref{app:gPP}).

In Fig.~\ref{fig:gpp} we show our calculated polariton-polariton interaction strength as a function of detuning for different values of the light-matter coupling.  We find that the simple approximation, $\beta^4 g_{\rm XX}$, works extremely well when $\Omega\ll\eb$, as expected.  Moreover, we see that $g_{\rm PP}$ shifts upwards compared to $\beta^4 g_{\rm XX}$ with increasing Rabi coupling, and the relative shift in the interaction strength is largest at negative detuning, where the polariton is photon dominated.  However, the difference between $g_{\rm PP}$ and $g^{(0)}_{\rm PP}$ is significantly smaller than that predicted by previous work~\cite{Tassone1999,Rochat2000,MCombescot2007,MCombescot2008,Glazov2009}.  Physically, this is because the light-matter coupling also decreases the size of the exciton in the LP, thus reducing the effect of exchange processes compared to the case where the exciton is taken to be rigid and unaffected by the light field.  Therefore, calculations that neglect light-induced changes to the electron-hole wave function will drastically overestimate the size of the correction to $g^{(0)}_{\rm PP}$ (see Appendix~\ref{app:gPP}).  Note that we never see a reduction of $g_{\rm PP}$ compared with $g^{(0)}_{\rm PP}$ as was predicted in Ref.~\cite{Xue2016} --- this is most likely because that paper used an unrenormalized theory.

\mmp{In obtaining the polariton-polariton interaction strength, we have made several approximations which are standard in the polariton literature (see, e.g., Ref.~\cite{Tassone1999}). Firstly, we have assumed that the system can be described by a 2D model, similarly to previous work on the electron-hole-photon model. Such an assumption may not be accurate in the case of GaAs quantum wells since the exciton size is comparable to the width of the quantum well, as discussed in Ref.~\cite{Estrecho2018}. However, it is an excellent approximation in the case of atomically thin materials~\cite{Wang2018}. Secondly, we employ the Born approximation, which does not capture the full energy dependence of exciton-exciton scattering. For instance, the low-energy scattering between neutral bosons in 2D should depend logarithmically on energy --- see, e.g., Refs.~\cite{Adhikari1986,2Dreview} --- rather than being constant. Furthermore, the exciton-exciton interaction should depend non-trivially on the electron-hole mass ratio~\cite{Petrov2005}. To improve upon the Born approximation, one must solve a complicated four-body problem which has so far only been achieved in the context of ultracold atoms, where the interactions are effectively contact~\cite{Petrov2003}. Thus, this goes far beyond the scope of the current work. However, we expect that our key result, that the light-induced change of the exciton significantly reduces the effect of exchange, will hold beyond our approximations.}

\section{Conclusions and outlook}

To summarize, we have presented a microscopic theory of exciton-polaritons in a semiconductor planar microcavity that explicitly involves electrons, holes, and photons. Crucially, we have shown that the light-matter coupling strongly renormalizes the cavity photon frequency in this model, a feature that has apparently been missed by all previous theoretical treatments.  The UV divergence of the photon self energy in our model is found to be akin to the vacuum polarization in quantum electrodynamics that acts to screen the electromagnetic field.  We have demonstrated how the redefinition of the exciton-photon detuning from its bare value leads to cutoff independent results, and how the microscopic model parameters can be related to the experimentally observed exciton-photon model parameters in the regime where $\Omega\ll\eb$. Our approach furthermore provides concrete predictions for the regime of very strong light-matter coupling $\Omega\sim\eb$, where the assumptions of the simple two-level model cease to be justified.

As we increase the Rabi coupling, we find that the upper polariton is strongly affected by its proximity to excited excitonic states and the unbound electron-hole continuum.  In particular, the weight of the upper polariton in the photon spectral response is visibly reduced as the photon becomes hybridized with higher energy electron-hole states.  By contrast, we find that the energy and photon fraction of the lower polariton are surprisingly well approximated by the two-level model in the very strong coupling regime.  However, the electron-hole wave functions of both upper and lower polaritons are significantly modified by the coupling to light, consistent with recent measurements~\cite{Brodbeck2017}. Our predictions for the photon spectral response and the electron-hole wave function can be tested in experiment.

Our work provides the foundation for future studies of few- and many-body physics within the microscopic model, since an understanding of the two-body electron-hole problem in the presence of light-matter coupling is essential to such treatments. As a first demonstration, we have used the exact electron-hole wave function to properly determine non-trivial light-induced corrections to the polariton-polariton interaction strength. Most notably, we have shown that the effect of particle exchange on the Rabi coupling is much smaller than previously thought~\cite{Tassone1999,Rochat2000,MCombescot2007,MCombescot2008,Glazov2009} due to the light-induced reduction of the exciton size in the lower polariton.  This result has important implications for ongoing measurements of the polariton-polariton interaction strength, since many experiments feature Rabi couplings that lie within the very strong coupling regime, \mmp{such as standard GaAs systems with multiple quantum wells~\cite{Estrecho2018}.}

Finally, we emphasize that our approach can be easily adapted to describe a range of other scenarios with strong light-matter coupling such as Rydberg exciton-polaritons in bulk materials~\cite{Rypolariton,Kazimierczuk2014,Heckotter2017} and excitonic resonances in an electron gas~\cite{Sidler2016}.


\acknowledgements

We gratefully acknowledge fruitful discussions with Francesca Maria Marchetti, Jonathan Keeling, Elena Ostrovskaya, Eliezer Estrecho, and Maciej Pieczarka.
We acknowledge support from the Australian Research Council Centre of Excellence in Future Low-Energy Electronics Technologies (CE170100039).
JL is furthermore supported through the Australian Research Council Future Fellowship FT160100244. 

 
\appendix

\section{Polaritons at finite momentum}\label{app:FiniteQ}

In this appendix, we derive the equations describing exciton-polaritons at finite momentum. The main difference from the scenario described in Sec.~\ref{sec:renorm} is that we now need to take the photon dispersion into account. However this does not change the renormalization procedure.

Similar to Eq.~\eqref{eq:psi}, we can write down the most general wave function with a finite center-of-mass momentum $\Q$:
\begin{equation}\label{eq:fullpsi}
\ket{\Psi^{(\Q)}}=\left( \sum_{\k} \varphi^{(\Q)}_\k e^\dagger_{\Qe+\k} h^\dagger_{\Qh-\k}+ \gamma^{(\Q)} c^{\dagger}_\Q\right) \ket{0},
\end{equation}
where $\varphi^{(\Q)}_\k$ is the electron-hole relative wave function and $\gamma^{(\Q)}$ represents the photon amplitude. Here, we find it convenient to define $\Q_{e,h}\equiv\frac{m_{e,h}}{m_e+m_h}\Q$ with $\Qe+\Qh=\Q$. By projecting the Schr\"{o}dinger equation $(E +\eg-\hat H)\ket{\Psi^{(\Q)}} = 0$ onto photon and electron-hole states, we arrive at the coupled equations:
\begin{subequations}\label{eq:finiteQ}
\begin{align}
\left[E-\omega_{X\Q}-\bar{\omega}_\k \right]\varphi^{(\Q)}_\k&=-\sum_{\kP} V(\k-\kP)\varphi^{(\Q)}_{\kP}+g \gamma^{(\Q)}\label{eq:reeq1}\\
\left[E-\omega_{c\Q}-\omega+\eg \right] \gamma^{(\Q)}&=g \sum_\k \varphi^{(\Q)}_\k, \label{eq:reeq2}
\end{align}
\end{subequations}
where $\omega_{X\Q}\equiv\frac{Q^2}{2 M}$ is the exciton dispersion with the mass $M\equiv m_e+m_h$. Eq.~\eqref{eq:finiteQ} is the finite-momentum analog of Eq.~\eqref{eq:vareqs} in Section~\ref{sec:renorm}.

To proceed, as in the Sec.~\ref{sec:renorm} we separate out the divergent part from the relative wave function $\varphi_\k^{(\Q)}$: %
\begin{align}
    \varphi_\k^{(\Q)}&=\beta_\k^{(\Q)}+\frac{g\gamma^{(\Q)}}{E-\omega_{X\Q}-\ok}.
\end{align}
Inserting this expression into Eq.~\eqref{eq:finiteQ} then yields two coupled equations for $\beta_\k^{(\Q)}$ and $\gamma^{(\Q)}$:
\begin{subequations}
\label{eq:betaapp}
\begin{align}
\label{eq:betaapp1}
&\left[E-\omega_{X\Q}-\ok\right]\beta_\k^{(\Q)} \nn \\ & \hspace{2mm}=-\sum_{\k'}V(\k-\k')\beta_{\k'}^{(\Q)}
+g\gamma^{(\Q)}\sum_{\k'}\frac{V(\k-\k')}{-E+\omega_{X\Q}+\bar\omega_{\k'}},\\
&\left[E-\omega_{c\Q}-\omega+\eg +g^2 \sum_\k \frac{1}{-E+\omega_{X\Q}+\bar{\omega}_\k} \right] \gamma^{(\Q)} \nn \\ &\hspace{2mm} = g\sum_\k \beta^{(\Q)}_\k.
\label{eq:betaapp2}
\end{align} 
\end{subequations}
Again, all sums converge except for the sum on the left hand side of Eq.~\eqref{eq:betaapp2}. Thus, we must again introduce the physical cavity photon frequency.

As in Sec.~\ref{sec:relation_to_two_level} we now consider the limit $g\ll a_0\eb$ where we can relate our model parameters to the usual two-level model used to describe exciton-polaritons. In this limit, the energies of interest satisfy $E\simeq -\eb+\omega_{X\Q}$, and we thus perform this replacement in the second sum on the right hand side of Eq.~\eqref{eq:betaapp1} and in the sum on the left hand side of Eq.~\eqref{eq:betaapp2}, analogously to how we used $E\simeq-\eb$ in Sec.~\ref{sec:relation_to_two_level}. Then the $\Q$ dependence becomes trivial and hence the renormalization procedure is exactly the same as in Sec.~\ref{sec:relation_to_two_level}. Thus, defining $\beta_\k^{(\Q)}\simeq\beta^{(\Q)}\tilde\Phi_{1s\k}$ and following the steps of Sec.~\ref{sec:relation_to_two_level}, the coupled set of equations~\eqref{eq:betaapp} become
\begin{align}
    \label{eq_energy_mat}
E\begin{pmatrix}
\beta^{(\Q)}\\
\gamma^{(\Q)}
\end{pmatrix}=\begin{pmatrix}
\omega_{X\Q}-\eb & \Omega\\
\Omega & \delta+\omega_{c\Q}-\eb
\end{pmatrix}\begin{pmatrix}
\beta^{(\Q)}\\
\gamma^{(\Q)}
\end{pmatrix}.
\end{align}
This results in the usual polariton dispersions given by the two-level model~\cite{Deng2010,Carusotto2013}:
\begin{align}\label{Eupdown}
E_{{\tiny  \begin{matrix} \text{LP}\\\text{UP}\end{matrix}}\Q}
=&-\eb+\frac{1}{2} (\omega_{X\Q}+\delta+\omega_{c\Q}) \nn \\ & \mp \frac{1}{2}\sqrt{\left(\omega_{X\Q}-\delta-\omega_{c\Q}\right)^2+4\Omega^2}.
\end{align}

The photon transmission as a function of center-of-mass momentum shown in Fig.~\ref{fig:transmission} is obtained by solving the coupled set of equations~\eqref{eq:finiteQ} numerically using the renormalized parameters.

\section{Photon spectrum and the 2D electron-hole $T$ matrix}\label{app:2DGreenfunction}

In this appendix, we provide further details on the diagrammatic approach to the renormalization of the cavity photon frequency presented in Sec.~\ref{sub:diag}. We also discuss the integral representation of the electron-hole $T$ matrix derived in Ref.~\cite{2DGreenfunction}.

Our starting point is the photon self energy shown in Fig.~\ref{fig:diagrams}(b) which we now write as
\begin{align}
    \Sigma(E) & = g^2\sum_{\k,\k'}G(\k,\k';E),
\label{eq:sigmaapp}
\end{align}
where $G(\k,\k';E)$ is the two-body electron-hole Green's function in the absence of light-matter coupling, with $\k$ and $\k'$ the incoming and outgoing relative momenta, respectively. This in turn satisfies the Lippmann-Schwinger equation~\cite{fetterbook}
\begin{align}
    G(\k,\k';E) & = \frac{\delta_{\k,\k'}}{E-\ok}-\sum_{\p}\frac{V(\k-\p)G(\p,\k';E)}{E-\bar\omega_\p}
    \label{eq:Gapp1}
\end{align}
(we again implicitly include a small positive imaginary part in the energy $E$). The solution of such an equation is commonly written in terms of a $T$ matrix 
\begin{align}
    G(\k,\k';E) & = \frac{\delta_{\k,\k'}}{E-\ok}+\frac{T(\k,\k';E)}{(E-\ok)(E-\bar\omega_{\k'})},
    \label{eq:Gapp2}
\end{align}
where $T(\k,\k';E)$ satisfies
\begin{align}
T(\k,\k';E)=-V(\k-\k')-\sum_{\p}\frac{V(\k-\p)T(\p,\k';E)}{E-\bar\omega_{\p}},
\label{eq:Tmat}
\end{align}
as illustrated in Fig.~\ref{fig:diagrams}(c).  Combining Eqs.~\eqref{eq:sigmaapp} and \eqref{eq:Gapp2} we see that we reproduce the expression for the self energy in Eq.~\eqref{eq:Sigma}.

We now show that the spectrum of the dressed photon propagator, Eq.~\eqref{eq:GC}, equals that obtained from the variational approach via the set of equations~\eqref{eq:betaeqs}. To see this, compare Eqs.~\eqref{eq:iib} and \eqref{eq:photonpole}. These equations have exactly the same functional form, provided we identify
\begin{align}
g\gamma \sum_{\k'}\frac{T(\k,\k';E)}{E-\bar\omega_{\k'}} &\equiv
(E-\bar\omega_\k)\beta_\k.
\label{eq:betaT}
\end{align}
Applying the operator $g\gamma\sum_{\k'}\frac1{E-\omega_{\k'}}\left\{\cdot\right\}$ to the equation satisfied by $T$, Eq.~\eqref{eq:Tmat}, using the symmetry of $T(\k,\k';E)$ with respect to $\k\leftrightarrow\k'$~\footnote{The symmetry of $T(\k,\k';E)$ is straightforward to see by iterating Eq.~\eqref{eq:Tmat} once.}, and using the replacement \eqref{eq:betaT}, we indeed find that we exactly reproduce the equation for $\beta_\k$, Eq.~\eqref{eq:ib}. Hence the left and right hand sides of Eq.~\eqref{eq:betaT} satisfy the exact same integral equation, and consequently the two approaches produce the same spectrum.

We also note that Eq.~\eqref{eq:betaT} implies that $\beta(0)=\sum_\k\beta_\k$ is finite as the momentum cutoff $\Lambda\to\infty$; this can be seen by simple momentum power counting of all terms produced by iterating Eq.~\eqref{eq:Tmat} in Eq.~\eqref{eq:betaT}. Thus, the function $\beta(r)$ is regular at the origin.

We now describe the integral representation of the 2D Coulomb Green's function derived in Ref.~\cite{2DGreenfunction}, and the resulting expression for the electron-hole $T$ matrix. The key to this representation is the mapping of the momentum $\k=(k_x,k_y)$ (and likewise $\k'$) in the equation satisfied by the Green's function, Eq.~\eqref{eq:Gapp1}, onto a three-dimensional (3D) unit sphere: Setting $E\equiv-k_0^2/2m_r$, we define the 3D unit vector 
\begin{align}
    \boldsymbol{\xi}&\equiv\frac1{\lambda(k)}\begin{pmatrix}2 k_0 k_x\\
2 k_0 k_y\\
k_0^2-k^2\end{pmatrix}
\end{align}
with $\lambda(k)\equiv k_0^2+k^2$. Taking now the continuum limit, we denote the elementary solid angle on the unit sphere as $\diff\boldsymbol{\xi}$.
Then we identify
\begin{subequations}
	\begin{align}
		&\diff\boldsymbol{\xi}=\left( \frac{2k_0}{\lambda(k)} \right)^2\diff \k\\
		&\delta(\boldsymbol{\xi}-\boldsymbol{\xi}^\prime)=\left( \frac{\lambda(k)}{2k_0} \right)^2 \delta(\k-\k')\\
		&|\boldsymbol{\xi}-\boldsymbol{\xi}'|^2=\frac{4 k_0^2}{\lambda(k)\lambda(k')}|\k-\k'|^2
	\end{align}
\end{subequations}
Under this mapping, the Green's function can be expressed by variables on the unit sphere as
\begin{align}\label{GtoGamma}
G(\k,\k';E)=-8 m_r k_0^2\frac{1}{\lambda(k)^{3/2}}\Gamma(\boldsymbol{\xi},\boldsymbol{\xi}^\prime)\frac{1}{\lambda(k')^{3/2}}.
\end{align}
Solving (the continuum limit of) Eq.~\eqref{eq:Gapp1} for $\Gamma(\boldsymbol{\xi},\boldsymbol{\xi}^\prime)$ yields~\cite{2DGreenfunction,2DGreencorrect}:
\begin{align}\label{Gamma_solusion}
\Gamma(\boldsymbol{\xi},\boldsymbol{\xi}^\prime)=&\delta(\boldsymbol{\xi}-\boldsymbol{\xi}^\prime)+\frac{\nu}{2\pi}\frac{1}{\left|\boldsymbol{\xi}-\boldsymbol{\xi}'\right|}\nn \\
&+\frac{\nu^2}{2\pi}\int_0^1 \diff u \frac{u^{-(\nu+1/2)}}{\left[(1-u)^2+u|\boldsymbol{\xi}-\boldsymbol{\xi}'|^2\right]^{1/2}},
\end{align}
where $\nu\equiv -\frac{1}{2 a_0}\frac{1}{k_0}$ (note that $a_0=k_0^{-1}$ for the $1s$ exciton ground state). Eq.~\eqref{GtoGamma} and Eq.~\eqref{Gamma_solusion} combined yield the two-body electron-hole Green’s function $G(\k,\k^\prime;E)$.

Finally, from the relationship between the electron-hole Green's function and the $T$ matrix, Eq.~\eqref{eq:Gapp2}, we find the integral representation of the two-body electron-hole scattering $T$ matrix in a manner similar to how this was recently done in three dimensions~\cite{3DCoulombTmatrix}:
\begin{align}\label{T2simeq}
T(\k,\k';E)=-\frac{k_0\nu z}{4\pi m_r \left|\k-\k'\right|}\int_0^1\frac{u^{-(\nu+1/2)}(1-u^2)}{\left[u+z(1-u)^2\right]^{3/2}}\diff u,
\end{align}
where $z\equiv \frac{\lambda(k)\lambda(k')}{4k_0^2|\k-\k'|^2}$.

\section{Polariton-polariton interactions \label{app:gPP}}

In this section, we discuss how to obtain the elastic interaction constant between lower polaritons, Eq.~\eqref{eq:gpp2}, from Eq.~\eqref{eq:gpp}: $g_{\rm{PP}}/A  = \frac12\bra{PP} \hat H-2E_g-2E_{\rm{LP}}\ket{PP}$. Using the explicit form of the polariton operator, Eq.~\eqref{eq:pol}, and performing all contractions between operators, we obtain
\begin{widetext}
\begin{align}
    \frac12\bra{PP} \hat H-2E_g\ket{PP} = & \ 2(\omega-\eg) \gamma_{\rm LP}^4+4\gamma_{\rm LP}^3g\sum_\k\varphi_{\rm{LP}\k}+2\gamma_{\rm LP}^2\sum_\k\varphi_{\rm{LP}\k}\left[\varphi_{\rm{LP}\k}(\omega-\eg+\bar\omega_\k)-\sum_{\k'}\varphi_{\rm{LP}\k'}V(\k-\k')\right]\nn \\ &-4\gamma_{\rm LP} g\sum_\k\varphi_{\rm{LP}\k}^2\left[\varphi_{\rm{LP}\k}-\sum_{\k'}\varphi_{\rm{LP}\k'}\right] -2\sum_{\k_1\k_2\k_3}\varphi_{\rm{LP}\k_1}^2\varphi_{\rm{LP}\k_2}\varphi_{\rm{LP}\k_3}V(\k_2-\k_3)\nn \\ &+4\sum_{\k_1\k_2}\varphi_{\rm{LP}\k_1}^3\varphi_{\rm{LP}\k_2}V(\k_1-\k_2)-2\sum_{\k_1\k_2}\varphi_{\rm{LP}\k_1}^2\varphi_{\rm{LP}\k_2}^2V(\k_1-\k_2)\nn \\ &+2\sum_{\k_1\k_2}\varphi_{\rm{LP}\k_1}^2\varphi_{\rm{LP}\k_2}^2\bar\omega_{\k_1}-2\sum_\k\varphi_{{\rm LP}\k}^4\bar\omega_\k.
    \label{eq:Hexpt}
\end{align}
\end{widetext}
Likewise, we find
\begin{align}
        \frac12\braket{PP}{PP}=1-\sum_\k\varphi_{\rm{LP}\k}^4.
        \label{eq:gppnorm}
\end{align}

At this point, a remark on units is in order. The interaction energy shift that we have calculated to extract $g_{\rm PP}$ is formally within an area $A$, which we have set to 1 in this work. Reinstating briefly this factor, the two terms of Eq.~\eqref{eq:gppnorm} are 1 and $O(a_0^2/A)$, respectively. However, the leading contribution (which does not scale as a scattering term) will cancel the corresponding contribution from Eq.~\eqref{eq:Hexpt}, such that in the end $g_{\rm PP}\sim \eb a_0^2$, as required.

Now, to arrive at Eq.~\eqref{eq:gpp2} in the main text, we apply the normalization of the wave function,
\begin{align}
        \gamma_{\rm LP}^2+\sum_\k\varphi_{\rm{LP}\k}^2=1,
\end{align}
as well as the equations satisfied by $\varphi_{\rm{LP}\k}$ and $\gamma_{\rm LP}$,
\begin{subequations}
\begin{align}
\label{eq:LPSE}
        (E_{\rm LP}-\ok)\varphi_{{\rm LP}\k} &=-\sum_{\k'}V(\k-\k')\varphi_{{\rm LP}\k'} +g\gamma_{\rm LP}, \\
        (E_{\rm LP}-\omega+\eg)\gamma_{\rm LP} & = g\sum_\k \varphi_{{\rm LP}\k}.
\end{align}
\end{subequations}
This latter equation is used repeatedly to systematically replace terms containing $\gamma_{\rm LP}$ with equivalent terms of a smaller power.

In the following two subsections, we discuss the limiting case of exciton-exciton scattering, and the previously considered perturbative corrections to this in the presence of strong light-matter coupling.

\subsection{Exciton-exciton scattering}
For completeness, we now discuss the evaluation of the interaction constant $g_{\rm XX}$ between two $1s$ excitons. This has been done previously in the literature, see e.g., Refs.~\cite{Tassone1999,Glazov2009}. As shown in Ref.~\cite{Tassone1999}, $g_{\rm XX}$ takes the form
\begin{align}
    g_{\rm{XX}} = 2 \sum_{\k,\k'} V(\k\!-\!\k')  \tilde\Phi_{1s\k}^3 \tilde\Phi_{1s\k'} \!-\! 2 \sum_{\k,\k'} V(\k\!-\!\k')  \tilde\Phi_{1s\k}^2  \tilde\Phi_{1s\k'}^2,
\end{align}
A simple application of the \sch equation satisfied by $ \tilde\Phi_{1s\k}$ shows that this can be written in the more symmetric form
\begin{align}
    g_{\rm{XX}} = 2 \sum_\k \left(\ok \! +\! \eb \right)  \tilde\Phi_{1s\k}^4 \!-\! 2 \sum_{\k,\k'} V(\k\!-\!\k')  \tilde\Phi_{1s\k}^2  \tilde\Phi_{1s\k'}^2,
\label{eq:gxxapp}
\end{align}
which is identical to Eq.~\eqref{eq:gpp2} when one takes $\varphi_{\rm{LP}\k}\to \tilde\Phi_{1s\k}$ and $E_{\rm LP}\to -\eb$. Using the $1s$ exciton wave function, Eq.~\eqref{eq:phi1s}, and the method of residues, the first term is straightforward to evaluate, with the result $\frac{8\pi}{2m_r}$.

The second term in Eq.~\eqref{eq:gxxapp} can be evaluated with great accuracy using the following trick to remove the integrable singularity originating from the Coulomb interaction. Start by shifting $\k-\k'\to\p$ and $\k+\k'\to2\q$. Then, the integral becomes
\begin{align}
    2\sum_{\p,\q}V(\p)\varphi^2_{{\rm LP},\frac{\p}2+\q}\varphi^2_{{\rm LP},\frac{\p}2-\q}.
\end{align}
This effectively removes the simple pole of the Coulomb interaction. We can now analytically perform the integral over first the angle between $\p$ and $\q$ and then over one of the momenta. The remaining integral is very well behaved numerically, and we find
\begin{align}
    2 \sum_{\k,\k'} V(\k\!-\!\k')  \tilde\Phi_{1s\k}^2  \tilde\Phi_{1s\k'}^2\simeq \frac{19.0761}{2m_r}.
\end{align}
Thus, in total we have
\begin{align}
g_{\rm XX}\simeq \frac{6.0566}{2m_r}=6.0566\eb a_0^2.
\end{align}
This is, of course, consistent with the results of Refs.~\cite{Tassone1999,Glazov2009}.

\subsection{Perturbative corrections to the polariton-polariton interaction}
Here we compare our microscopic results with perturbative corrections to $g_{\rm PP}^{(0)} = \beta^4g_{XX}$ previously considered in the literature.  The first of these is due to exciton oscillator saturation as investigated by Tassone and Yamamoto in Ref.~\cite{Tassone1999} (see also Refs.~\cite{Rochat2000,Glazov2009,Brichkin2011}). Secondly, we can consider the photon assisted exchange processes due to the strong off-shell scattering of excitons when coupled to light.  We stress that both of these corrections assume that the relative electron-hole wave function in the lower polariton is unchanged from the $1s$ exciton state, unlike what we find in our microscopic model.

The exciton oscillator strength saturation~\cite{Tassone1999} has been estimated to lead to the correction~\cite{Brichkin2011}
\begin{align}
    \Delta g_{\rm PP}^{\rm sat}=\frac{32\pi}{7}\Omega a_0^2\beta^3|\gamma|.
    \label{eq:corra}
\end{align}
This is expected to be important in the very strong coupling regime when $\Omega\sim\eb$~\cite{Tassone1999}. We can see how such a term arises naturally within our microscopic calculation, although we disagree with the prefactor. Using the \sch equation satisfied by the lower polariton, Eq.~\eqref{eq:LPSE}, to replace $E_{\rm LP}\varphi_{{\rm LP}\k}$ in Eq.~\eqref{eq:gpp2}, and taking $\varphi_{\rm LP\k}\to\beta\tilde\Phi_{1s\k}$ we find the correction to $g_{\rm PP}^{(0)}$ to be:
\begin{align}
    2g|\gamma|\beta^3\sum_\k \tilde \Phi_{1s\k}^3=\frac{16\pi}7\Omega a_0^2\beta^3|\gamma|.
\end{align}
We note that $\frac{1}2 \bra{PP} \hat H\ket{PP}$ alone contains the term $\frac{32\pi}7\Omega a_0^2\beta^3|\gamma|$. However, subtracting the normalization as in Eq.~\eqref{eq:gpp} reduces this contribution to $\frac{16\pi}7\Omega a_0^2\beta^3|\gamma|$. This may explain the discrepancy in prefactors.

On the other hand, we can also consider the correction one would obtain if we take into account the change in the collision energy due to the photon detuning, while keeping the wave function unchanged. Under this approximation, we would have
\begin{align}
    \Delta g_{\rm PP}^{\Delta E}&=-(E_{{\rm LP}}+\eb)\beta^4\sum_\k \tilde\Phi_{1s\k}^4 \nn \\ 
    &\simeq \frac12\left(\sqrt{\delta^2+4\Omega^2}-\delta\right)\frac{16\pi a_0^2\beta^4}{5},
    \label{eq:corrb}
\end{align}
where in the last step we used the two-level expression for the lower polariton energy, Eq.~\eqref{eq:two-level dispersion}. Eq.~\eqref{eq:corrb} has a form similar to the photon assisted exchange scattering matrix element derived in Ref.~\cite{MCombescot2007} (see also \cite{Betbeder-Matibet2002}).

As we have shown in Fig.~\ref{fig:gpp}, when $\Omega\ll\eb$ the simple exciton approximation $g_{\rm PP}\simeq g_{\rm PP}^{(0)}\equiv \beta^4g_{\rm XX}$ works very well. In the limit of very strong light-matter coupling where we find appreciable corrections to this result, the two perturbative corrections in Eqs.~\eqref{eq:corra} and \eqref{eq:corrb} greatly overestimate these. For instance, at zero detuning and $\Omega/\eb=0.5$ we find $g_{\rm PP}-g_{\rm PP}^{(0)}\simeq 0.25\eb a_0^2$, whereas Eq.~\eqref{eq:corra} predicts a shift of $1.8\eb a_0^2$ and \eqref{eq:corrb} a shift of $1.26\eb a_0^2$.

\bibliography{polariton_refs}

\begin{thebibliography}{66}%
\makeatletter
\providecommand \@ifxundefined [1]{%
 \@ifx{#1\undefined}
}%
\providecommand \@ifnum [1]{%
 \ifnum #1\expandafter \@firstoftwo
 \else \expandafter \@secondoftwo
 \fi
}%
\providecommand \@ifx [1]{%
 \ifx #1\expandafter \@firstoftwo
 \else \expandafter \@secondoftwo
 \fi
}%
\providecommand \natexlab [1]{#1}%
\providecommand \enquote  [1]{``#1''}%
\providecommand \bibnamefont  [1]{#1}%
\providecommand \bibfnamefont [1]{#1}%
\providecommand \citenamefont [1]{#1}%
\providecommand \href@noop [0]{\@secondoftwo}%
\providecommand \href [0]{\begingroup \@sanitize@url \@href}%
\providecommand \@href[1]{\@@startlink{#1}\@@href}%
\providecommand \@@href[1]{\endgroup#1\@@endlink}%
\providecommand \@sanitize@url [0]{\catcode `\\12\catcode `\$12\catcode
  `\&12\catcode `\#12\catcode `\^12\catcode `\_12\catcode `\%12\relax}%
\providecommand \@@startlink[1]{}%
\providecommand \@@endlink[0]{}%
\providecommand \url  [0]{\begingroup\@sanitize@url \@url }%
\providecommand \@url [1]{\endgroup\@href {#1}{\urlprefix }}%
\providecommand \urlprefix  [0]{URL }%
\providecommand \Eprint [0]{\href }%
\providecommand \doibase [0]{http://dx.doi.org/}%
\providecommand \selectlanguage [0]{\@gobble}%
\providecommand \bibinfo  [0]{\@secondoftwo}%
\providecommand \bibfield  [0]{\@secondoftwo}%
\providecommand \translation [1]{[#1]}%
\providecommand \BibitemOpen [0]{}%
\providecommand \bibitemStop [0]{}%
\providecommand \bibitemNoStop [0]{.\EOS\space}%
\providecommand \EOS [0]{\spacefactor3000\relax}%
\providecommand \BibitemShut  [1]{\csname bibitem#1\endcsname}%
\let\auto@bib@innerbib\@empty
\bibitem [{\citenamefont {Kavokin}\ \emph {et~al.}(2017)\citenamefont
  {Kavokin}, \citenamefont {Baumberg}, \citenamefont {Malpuech},\ and\
  \citenamefont {Laussy}}]{Microcavities}%
  \BibitemOpen
  \bibfield  {author} {\bibinfo {author} {\bibfnamefont {A.~V.}\ \bibnamefont
  {Kavokin}}, \bibinfo {author} {\bibfnamefont {J.~J.}\ \bibnamefont
  {Baumberg}}, \bibinfo {author} {\bibfnamefont {G.}~\bibnamefont {Malpuech}},
  \ and\ \bibinfo {author} {\bibfnamefont {F.~P.}\ \bibnamefont {Laussy}},\
  }\href
  {https://global.oup.com/academic/product/microcavities-9780198782995?cc=au&lang=en&#}
  {\emph {\bibinfo {title} {Microcavities}}}\ (\bibinfo  {publisher} {Oxford
  University Press},\ \bibinfo {year} {2017})\BibitemShut {NoStop}%
\bibitem [{\citenamefont {Keeling}\ \emph {et~al.}(2007)\citenamefont
  {Keeling}, \citenamefont {Marchetti}, \citenamefont {Szyma{\'n}ska},\ and\
  \citenamefont {Littlewood}}]{Keeling2007}%
  \BibitemOpen
  \bibfield  {author} {\bibinfo {author} {\bibfnamefont {J.}~\bibnamefont
  {Keeling}}, \bibinfo {author} {\bibfnamefont {F.~M.}\ \bibnamefont
  {Marchetti}}, \bibinfo {author} {\bibfnamefont {M.~H.}\ \bibnamefont
  {Szyma{\'n}ska}}, \ and\ \bibinfo {author} {\bibfnamefont {P.~B.}\
  \bibnamefont {Littlewood}},\ }\bibfield  {title} {\bibinfo {title} {\emph
  {Collective coherence in planar semiconductor microcavities}},\ }\href
  {http://stacks.iop.org/0268-1242/22/i=5/a=R01} {\bibfield  {journal}
  {\bibinfo  {journal} {Semiconductor Science and Technology}\ }\textbf
  {\bibinfo {volume} {22}},\ \bibinfo {pages} {R1} (\bibinfo {year}
  {2007})}\BibitemShut {NoStop}%
\bibitem [{\citenamefont {Deng}\ \emph {et~al.}(2010)\citenamefont {Deng},
  \citenamefont {Haug},\ and\ \citenamefont {Yamamoto}}]{Deng2010}%
  \BibitemOpen
  \bibfield  {author} {\bibinfo {author} {\bibfnamefont {H.}~\bibnamefont
  {Deng}}, \bibinfo {author} {\bibfnamefont {H.}~\bibnamefont {Haug}}, \ and\
  \bibinfo {author} {\bibfnamefont {Y.}~\bibnamefont {Yamamoto}},\ }\bibfield
  {title} {\bibinfo {title} {\emph {Exciton-polariton Bose-Einstein
  condensation}},\ }\href {\doibase 10.1103/RevModPhys.82.1489} {\bibfield
  {journal} {\bibinfo  {journal} {Rev. Mod. Phys.}\ }\textbf {\bibinfo {volume}
  {82}},\ \bibinfo {pages} {1489} (\bibinfo {year} {2010})}\BibitemShut
  {NoStop}%
\bibitem [{\citenamefont {Carusotto}\ and\ \citenamefont
  {Ciuti}(2013)}]{Carusotto2013}%
  \BibitemOpen
  \bibfield  {author} {\bibinfo {author} {\bibfnamefont {I.}~\bibnamefont
  {Carusotto}}\ and\ \bibinfo {author} {\bibfnamefont {C.}~\bibnamefont
  {Ciuti}},\ }\bibfield  {title} {\bibinfo {title} {\emph {Quantum fluids of
  light}},\ }\href {\doibase 10.1103/RevModPhys.85.299} {\bibfield  {journal}
  {\bibinfo  {journal} {Rev. Mod. Phys.}\ }\textbf {\bibinfo {volume} {85}},\
  \bibinfo {pages} {299} (\bibinfo {year} {2013})}\BibitemShut {NoStop}%
\bibitem [{\citenamefont {Kasprzak}\ \emph {et~al.}(2006)\citenamefont
  {Kasprzak}, \citenamefont {Richard}, \citenamefont {Kundermann},
  \citenamefont {Baas}, \citenamefont {Jeambrun}, \citenamefont {Keeling},
  \citenamefont {Marchetti}, \citenamefont {Szyma{\'n}ska}, \citenamefont
  {Andr{\'{e}}}, \citenamefont {Staehli}, \citenamefont {Savona}, \citenamefont
  {Littlewood}, \citenamefont {Deveaud},\ and\ \citenamefont
  {Dang}}]{Kasprzak2006}%
  \BibitemOpen
  \bibfield  {author} {\bibinfo {author} {\bibfnamefont {J.}~\bibnamefont
  {Kasprzak}}, \bibinfo {author} {\bibfnamefont {M.}~\bibnamefont {Richard}},
  \bibinfo {author} {\bibfnamefont {S.}~\bibnamefont {Kundermann}}, \bibinfo
  {author} {\bibfnamefont {A.}~\bibnamefont {Baas}}, \bibinfo {author}
  {\bibfnamefont {P.}~\bibnamefont {Jeambrun}}, \bibinfo {author}
  {\bibfnamefont {J.~M.~J.}\ \bibnamefont {Keeling}}, \bibinfo {author}
  {\bibfnamefont {F.~M.}\ \bibnamefont {Marchetti}}, \bibinfo {author}
  {\bibfnamefont {M.~H.}\ \bibnamefont {Szyma{\'n}ska}}, \bibinfo {author}
  {\bibfnamefont {R.}~\bibnamefont {Andr{\'{e}}}}, \bibinfo {author}
  {\bibfnamefont {J.~L.}\ \bibnamefont {Staehli}}, \bibinfo {author}
  {\bibfnamefont {V.}~\bibnamefont {Savona}}, \bibinfo {author} {\bibfnamefont
  {P.~B.}\ \bibnamefont {Littlewood}}, \bibinfo {author} {\bibfnamefont
  {B.}~\bibnamefont {Deveaud}}, \ and\ \bibinfo {author} {\bibfnamefont
  {L.~S.}\ \bibnamefont {Dang}},\ }\bibfield  {title} {\bibinfo {title} {\emph
  {{Bose-Einstein condensation of exciton polaritons.}}},\ }\href {\doibase
  10.1038/nature05131} {\bibfield  {journal} {\bibinfo  {journal} {Nature}\
  }\textbf {\bibinfo {volume} {443}},\ \bibinfo {pages} {409} (\bibinfo {year}
  {2006})}\BibitemShut {NoStop}%
\bibitem [{\citenamefont {Balili}\ \emph {et~al.}(2007)\citenamefont {Balili},
  \citenamefont {Hartwell}, \citenamefont {Snoke}, \citenamefont {Pfeiffer},\
  and\ \citenamefont {West}}]{Balili1007}%
  \BibitemOpen
  \bibfield  {author} {\bibinfo {author} {\bibfnamefont {R.}~\bibnamefont
  {Balili}}, \bibinfo {author} {\bibfnamefont {V.}~\bibnamefont {Hartwell}},
  \bibinfo {author} {\bibfnamefont {D.}~\bibnamefont {Snoke}}, \bibinfo
  {author} {\bibfnamefont {L.}~\bibnamefont {Pfeiffer}}, \ and\ \bibinfo
  {author} {\bibfnamefont {K.}~\bibnamefont {West}},\ }\bibfield  {title}
  {\bibinfo {title} {\emph {Bose-Einstein Condensation of Microcavity
  Polaritons in a Trap}},\ }\href {\doibase 10.1126/science.1140990} {\bibfield
   {journal} {\bibinfo  {journal} {Science}\ }\textbf {\bibinfo {volume}
  {316}},\ \bibinfo {pages} {1007} (\bibinfo {year} {2007})}\BibitemShut
  {NoStop}%
\bibitem [{\citenamefont {Utsunomiya}\ \emph {et~al.}(2008)\citenamefont
  {Utsunomiya}, \citenamefont {Tian}, \citenamefont {Roumpos}, \citenamefont
  {Lai}, \citenamefont {Kumada}, \citenamefont {Fujisawa}, \citenamefont
  {Kuwata-Gonokami}, \citenamefont {L{\"o}ffler}, \citenamefont {H{\"o}fling},
  \citenamefont {Forchel},\ and\ \citenamefont {Yamamoto}}]{Utsunomiya2008}%
  \BibitemOpen
  \bibfield  {author} {\bibinfo {author} {\bibfnamefont {S.}~\bibnamefont
  {Utsunomiya}}, \bibinfo {author} {\bibfnamefont {L.}~\bibnamefont {Tian}},
  \bibinfo {author} {\bibfnamefont {G.}~\bibnamefont {Roumpos}}, \bibinfo
  {author} {\bibfnamefont {C.~W.}\ \bibnamefont {Lai}}, \bibinfo {author}
  {\bibfnamefont {N.}~\bibnamefont {Kumada}}, \bibinfo {author} {\bibfnamefont
  {T.}~\bibnamefont {Fujisawa}}, \bibinfo {author} {\bibfnamefont
  {M.}~\bibnamefont {Kuwata-Gonokami}}, \bibinfo {author} {\bibfnamefont
  {A.}~\bibnamefont {L{\"o}ffler}}, \bibinfo {author} {\bibfnamefont
  {S.}~\bibnamefont {H{\"o}fling}}, \bibinfo {author} {\bibfnamefont
  {A.}~\bibnamefont {Forchel}}, \ and\ \bibinfo {author} {\bibfnamefont
  {Y.}~\bibnamefont {Yamamoto}},\ }\bibfield  {title} {\bibinfo {title} {\emph
  {Observation of Bogoliubov excitations in exciton-polariton condensates}},\
  }\href {https://doi.org/10.1038/nphys1034} {\bibfield  {journal} {\bibinfo
  {journal} {Nature Physics}\ }\textbf {\bibinfo {volume} {4}},\ \bibinfo
  {pages} {700} (\bibinfo {year} {2008})}\BibitemShut {NoStop}%
\bibitem [{\citenamefont {Kohnle}\ \emph {et~al.}(2011)\citenamefont {Kohnle},
  \citenamefont {L\'eger}, \citenamefont {Wouters}, \citenamefont {Richard},
  \citenamefont {Portella-Oberli},\ and\ \citenamefont
  {Deveaud-Pl\'edran}}]{KohnlePRL2011}%
  \BibitemOpen
  \bibfield  {author} {\bibinfo {author} {\bibfnamefont {V.}~\bibnamefont
  {Kohnle}}, \bibinfo {author} {\bibfnamefont {Y.}~\bibnamefont {L\'eger}},
  \bibinfo {author} {\bibfnamefont {M.}~\bibnamefont {Wouters}}, \bibinfo
  {author} {\bibfnamefont {M.}~\bibnamefont {Richard}}, \bibinfo {author}
  {\bibfnamefont {M.~T.}\ \bibnamefont {Portella-Oberli}}, \ and\ \bibinfo
  {author} {\bibfnamefont {B.}~\bibnamefont {Deveaud-Pl\'edran}},\ }\bibfield
  {title} {\bibinfo {title} {\emph {From Single Particle to Superfluid
  Excitations in a Dissipative Polariton Gas}},\ }\href {\doibase
  10.1103/PhysRevLett.106.255302} {\bibfield  {journal} {\bibinfo  {journal}
  {Phys. Rev. Lett.}\ }\textbf {\bibinfo {volume} {106}},\ \bibinfo {pages}
  {255302} (\bibinfo {year} {2011})}\BibitemShut {NoStop}%
\bibitem [{\citenamefont {Roumpos}\ \emph {et~al.}(2010)\citenamefont
  {Roumpos}, \citenamefont {Fraser}, \citenamefont {L{\"o}ffler}, \citenamefont
  {H{\"o}fling}, \citenamefont {Forchel},\ and\ \citenamefont
  {Yamamoto}}]{Roumpos2010}%
  \BibitemOpen
  \bibfield  {author} {\bibinfo {author} {\bibfnamefont {G.}~\bibnamefont
  {Roumpos}}, \bibinfo {author} {\bibfnamefont {M.~D.}\ \bibnamefont {Fraser}},
  \bibinfo {author} {\bibfnamefont {A.}~\bibnamefont {L{\"o}ffler}}, \bibinfo
  {author} {\bibfnamefont {S.}~\bibnamefont {H{\"o}fling}}, \bibinfo {author}
  {\bibfnamefont {A.}~\bibnamefont {Forchel}}, \ and\ \bibinfo {author}
  {\bibfnamefont {Y.}~\bibnamefont {Yamamoto}},\ }\bibfield  {title} {\bibinfo
  {title} {\emph {Single vortex-antivortex pair in an exciton-polariton
  condensate}},\ }\href {https://doi.org/10.1038/nphys1841} {\bibfield
  {journal} {\bibinfo  {journal} {Nature Physics}\ }\textbf {\bibinfo {volume}
  {7}},\ \bibinfo {pages} {129} (\bibinfo {year} {2010})}\BibitemShut {NoStop}%
\bibitem [{\citenamefont {Dall}\ \emph {et~al.}(2014)\citenamefont {Dall},
  \citenamefont {Fraser}, \citenamefont {Desyatnikov}, \citenamefont {Li},
  \citenamefont {Brodbeck}, \citenamefont {Kamp}, \citenamefont {Schneider},
  \citenamefont {H\"ofling},\ and\ \citenamefont {Ostrovskaya}}]{RobPRL14}%
  \BibitemOpen
  \bibfield  {author} {\bibinfo {author} {\bibfnamefont {R.}~\bibnamefont
  {Dall}}, \bibinfo {author} {\bibfnamefont {M.~D.}\ \bibnamefont {Fraser}},
  \bibinfo {author} {\bibfnamefont {A.~S.}\ \bibnamefont {Desyatnikov}},
  \bibinfo {author} {\bibfnamefont {G.}~\bibnamefont {Li}}, \bibinfo {author}
  {\bibfnamefont {S.}~\bibnamefont {Brodbeck}}, \bibinfo {author}
  {\bibfnamefont {M.}~\bibnamefont {Kamp}}, \bibinfo {author} {\bibfnamefont
  {C.}~\bibnamefont {Schneider}}, \bibinfo {author} {\bibfnamefont
  {S.}~\bibnamefont {H\"ofling}}, \ and\ \bibinfo {author} {\bibfnamefont
  {E.~A.}\ \bibnamefont {Ostrovskaya}},\ }\bibfield  {title} {\bibinfo {title}
  {\emph {Creation of Orbital Angular Momentum States with Chiral Polaritonic
  Lenses}},\ }\href {\doibase 10.1103/PhysRevLett.113.200404} {\bibfield
  {journal} {\bibinfo  {journal} {Phys. Rev. Lett.}\ }\textbf {\bibinfo
  {volume} {113}},\ \bibinfo {pages} {200404} (\bibinfo {year}
  {2014})}\BibitemShut {NoStop}%
\bibitem [{\citenamefont {Mu{\~n}oz-Matutano}\ \emph
  {et~al.}(2019)\citenamefont {Mu{\~n}oz-Matutano}, \citenamefont {Wood},
  \citenamefont {Johnsson}, \citenamefont {Vidal}, \citenamefont {Baragiola},
  \citenamefont {Reinhard}, \citenamefont {Lema{\^\i}tre}, \citenamefont
  {Bloch}, \citenamefont {Amo}, \citenamefont {Nogues}, \citenamefont {Besga},
  \citenamefont {Richard},\ and\ \citenamefont {Volz}}]{Volz2017}%
  \BibitemOpen
  \bibfield  {author} {\bibinfo {author} {\bibfnamefont {G.}~\bibnamefont
  {Mu{\~n}oz-Matutano}}, \bibinfo {author} {\bibfnamefont {A.}~\bibnamefont
  {Wood}}, \bibinfo {author} {\bibfnamefont {M.}~\bibnamefont {Johnsson}},
  \bibinfo {author} {\bibfnamefont {X.}~\bibnamefont {Vidal}}, \bibinfo
  {author} {\bibfnamefont {B.~Q.}\ \bibnamefont {Baragiola}}, \bibinfo {author}
  {\bibfnamefont {A.}~\bibnamefont {Reinhard}}, \bibinfo {author}
  {\bibfnamefont {A.}~\bibnamefont {Lema{\^\i}tre}}, \bibinfo {author}
  {\bibfnamefont {J.}~\bibnamefont {Bloch}}, \bibinfo {author} {\bibfnamefont
  {A.}~\bibnamefont {Amo}}, \bibinfo {author} {\bibfnamefont {G.}~\bibnamefont
  {Nogues}}, \bibinfo {author} {\bibfnamefont {B.}~\bibnamefont {Besga}},
  \bibinfo {author} {\bibfnamefont {M.}~\bibnamefont {Richard}}, \ and\
  \bibinfo {author} {\bibfnamefont {T.}~\bibnamefont {Volz}},\ }\bibfield
  {title} {\bibinfo {title} {\emph {Emergence of quantum correlations from
  interacting fibre-cavity polaritons}},\ }\href {\doibase
  10.1038/s41563-019-0281-z} {\bibfield  {journal} {\bibinfo  {journal} {Nature
  Materials}\ }\textbf {\bibinfo {volume} {18}},\ \bibinfo {pages} {213}
  (\bibinfo {year} {2019})}\BibitemShut {NoStop}%
\bibitem [{\citenamefont {Delteil}\ \emph {et~al.}(2019)\citenamefont
  {Delteil}, \citenamefont {Fink}, \citenamefont {Schade}, \citenamefont
  {H{\"o}fling}, \citenamefont {Schneider},\ and\ \citenamefont {{\.I}mamo{\u
  g}lu}}]{Delteil2018}%
  \BibitemOpen
  \bibfield  {author} {\bibinfo {author} {\bibfnamefont {A.}~\bibnamefont
  {Delteil}}, \bibinfo {author} {\bibfnamefont {T.}~\bibnamefont {Fink}},
  \bibinfo {author} {\bibfnamefont {A.}~\bibnamefont {Schade}}, \bibinfo
  {author} {\bibfnamefont {S.}~\bibnamefont {H{\"o}fling}}, \bibinfo {author}
  {\bibfnamefont {C.}~\bibnamefont {Schneider}}, \ and\ \bibinfo {author}
  {\bibfnamefont {A.}~\bibnamefont {{\.I}mamo{\u g}lu}},\ }\bibfield  {title}
  {\bibinfo {title} {\emph {Towards polariton blockade of confined
  exciton--polaritons}},\ }\href {\doibase 10.1038/s41563-019-0282-y}
  {\bibfield  {journal} {\bibinfo  {journal} {Nature Materials}\ }\textbf
  {\bibinfo {volume} {18}},\ \bibinfo {pages} {219} (\bibinfo {year}
  {2019})}\BibitemShut {NoStop}%
\bibitem [{\citenamefont {Ciuti}\ \emph {et~al.}(1998)\citenamefont {Ciuti},
  \citenamefont {Savona}, \citenamefont {Piermarocchi}, \citenamefont
  {Quattropani},\ and\ \citenamefont {Schwendimann}}]{Ciuti1998}%
  \BibitemOpen
  \bibfield  {author} {\bibinfo {author} {\bibfnamefont {C.}~\bibnamefont
  {Ciuti}}, \bibinfo {author} {\bibfnamefont {V.}~\bibnamefont {Savona}},
  \bibinfo {author} {\bibfnamefont {C.}~\bibnamefont {Piermarocchi}}, \bibinfo
  {author} {\bibfnamefont {A.}~\bibnamefont {Quattropani}}, \ and\ \bibinfo
  {author} {\bibfnamefont {P.}~\bibnamefont {Schwendimann}},\ }\bibfield
  {title} {\bibinfo {title} {\emph {Role of the exchange of carriers in elastic
  exciton-exciton scattering in quantum wells}},\ }\href {\doibase
  10.1103/PhysRevB.58.7926} {\bibfield  {journal} {\bibinfo  {journal} {Phys.
  Rev. B}\ }\textbf {\bibinfo {volume} {58}},\ \bibinfo {pages} {7926}
  (\bibinfo {year} {1998})}\BibitemShut {NoStop}%
\bibitem [{\citenamefont {Tassone}\ and\ \citenamefont
  {Yamamoto}(1999)}]{Tassone1999}%
  \BibitemOpen
  \bibfield  {author} {\bibinfo {author} {\bibfnamefont {F.}~\bibnamefont
  {Tassone}}\ and\ \bibinfo {author} {\bibfnamefont {Y.}~\bibnamefont
  {Yamamoto}},\ }\bibfield  {title} {\bibinfo {title} {\emph {Exciton-exciton
  scattering dynamics in a semiconductor microcavity and stimulated scattering
  into polaritons}},\ }\href {\doibase 10.1103/PhysRevB.59.10830} {\bibfield
  {journal} {\bibinfo  {journal} {Phys. Rev. B}\ }\textbf {\bibinfo {volume}
  {59}},\ \bibinfo {pages} {10830} (\bibinfo {year} {1999})}\BibitemShut
  {NoStop}%
\bibitem [{\citenamefont {Rochat}\ \emph {et~al.}(2000)\citenamefont {Rochat},
  \citenamefont {Ciuti}, \citenamefont {Savona}, \citenamefont {Piermarocchi},
  \citenamefont {Quattropani},\ and\ \citenamefont
  {Schwendimann}}]{Rochat2000}%
  \BibitemOpen
  \bibfield  {author} {\bibinfo {author} {\bibfnamefont {G.}~\bibnamefont
  {Rochat}}, \bibinfo {author} {\bibfnamefont {C.}~\bibnamefont {Ciuti}},
  \bibinfo {author} {\bibfnamefont {V.}~\bibnamefont {Savona}}, \bibinfo
  {author} {\bibfnamefont {C.}~\bibnamefont {Piermarocchi}}, \bibinfo {author}
  {\bibfnamefont {A.}~\bibnamefont {Quattropani}}, \ and\ \bibinfo {author}
  {\bibfnamefont {P.}~\bibnamefont {Schwendimann}},\ }\bibfield  {title}
  {\bibinfo {title} {\emph {Excitonic Bloch equations for a two-dimensional
  system of interacting excitons}},\ }\href {\doibase
  10.1103/PhysRevB.61.13856} {\bibfield  {journal} {\bibinfo  {journal} {Phys.
  Rev. B}\ }\textbf {\bibinfo {volume} {61}},\ \bibinfo {pages} {13856}
  (\bibinfo {year} {2000})}\BibitemShut {NoStop}%
\bibitem [{\citenamefont {{Combescot, M.}}\ \emph {et~al.}(2007)\citenamefont
  {{Combescot, M.}}, \citenamefont {{Dupertuis, M. A.}},\ and\ \citenamefont
  {{Betbeder-Matibet, O.}}}]{MCombescot2007}%
  \BibitemOpen
  \bibfield  {author} {\bibinfo {author} {\bibnamefont {{Combescot, M.}}},
  \bibinfo {author} {\bibnamefont {{Dupertuis, M. A.}}}, \ and\ \bibinfo
  {author} {\bibnamefont {{Betbeder-Matibet, O.}}},\ }\bibfield  {title}
  {\bibinfo {title} {\emph {Polariton-polariton scattering: Exact results
  through a novel approach}},\ }\href {\doibase 10.1209/0295-5075/79/17001}
  {\bibfield  {journal} {\bibinfo  {journal} {EPL}\ }\textbf {\bibinfo {volume}
  {79}},\ \bibinfo {pages} {17001} (\bibinfo {year} {2007})}\BibitemShut
  {NoStop}%
\bibitem [{\citenamefont {Xue}\ \emph {et~al.}(2016)\citenamefont {Xue},
  \citenamefont {Wu}, \citenamefont {Xie}, \citenamefont {Su},\ and\
  \citenamefont {MacDonald}}]{Xue2016}%
  \BibitemOpen
  \bibfield  {author} {\bibinfo {author} {\bibfnamefont {F.}~\bibnamefont
  {Xue}}, \bibinfo {author} {\bibfnamefont {F.}~\bibnamefont {Wu}}, \bibinfo
  {author} {\bibfnamefont {M.}~\bibnamefont {Xie}}, \bibinfo {author}
  {\bibfnamefont {J.-J.}\ \bibnamefont {Su}}, \ and\ \bibinfo {author}
  {\bibfnamefont {A.~H.}\ \bibnamefont {MacDonald}},\ }\bibfield  {title}
  {\bibinfo {title} {\emph {Microscopic theory of equilibrium polariton
  condensates}},\ }\href {\doibase 10.1103/PhysRevB.94.235302} {\bibfield
  {journal} {\bibinfo  {journal} {Phys. Rev. B}\ }\textbf {\bibinfo {volume}
  {94}},\ \bibinfo {pages} {235302} (\bibinfo {year} {2016})}\BibitemShut
  {NoStop}%
\bibitem [{\citenamefont {Khurgin}(2001)}]{Khurgin2001}%
  \BibitemOpen
  \bibfield  {author} {\bibinfo {author} {\bibfnamefont {J.}~\bibnamefont
  {Khurgin}},\ }\bibfield  {title} {\bibinfo {title} {\emph {Excitonic radius
  in the cavity polariton in the regime of very strong coupling}},\ }\href
  {\doibase https://doi.org/10.1016/S0038-1098(00)00469-5} {\bibfield
  {journal} {\bibinfo  {journal} {Solid State Communications}\ }\textbf
  {\bibinfo {volume} {117}},\ \bibinfo {pages} {307 } (\bibinfo {year}
  {2001})}\BibitemShut {NoStop}%
\bibitem [{\citenamefont {Mak}\ and\ \citenamefont {Shan}(2016)}]{Mak2016}%
  \BibitemOpen
  \bibfield  {author} {\bibinfo {author} {\bibfnamefont {K.~F.}\ \bibnamefont
  {Mak}}\ and\ \bibinfo {author} {\bibfnamefont {J.}~\bibnamefont {Shan}},\
  }\bibfield  {title} {\bibinfo {title} {\emph {Photonics and optoelectronics
  of 2D semiconductor transition metal dichalcogenides}},\ }\href
  {https://doi.org/10.1038/nphoton.2015.282} {\bibfield  {journal} {\bibinfo
  {journal} {Nature Photonics}\ }\textbf {\bibinfo {volume} {10}},\ \bibinfo
  {pages} {216} (\bibinfo {year} {2016})}\BibitemShut {NoStop}%
\bibitem [{\citenamefont {Low}\ \emph {et~al.}(2016)\citenamefont {Low},
  \citenamefont {Chaves}, \citenamefont {Caldwell}, \citenamefont {Kumar},
  \citenamefont {Fang}, \citenamefont {Avouris}, \citenamefont {Heinz},
  \citenamefont {Guinea}, \citenamefont {Martin-Moreno},\ and\ \citenamefont
  {Koppens}}]{Low2016}%
  \BibitemOpen
  \bibfield  {author} {\bibinfo {author} {\bibfnamefont {T.}~\bibnamefont
  {Low}}, \bibinfo {author} {\bibfnamefont {A.}~\bibnamefont {Chaves}},
  \bibinfo {author} {\bibfnamefont {J.~D.}\ \bibnamefont {Caldwell}}, \bibinfo
  {author} {\bibfnamefont {A.}~\bibnamefont {Kumar}}, \bibinfo {author}
  {\bibfnamefont {N.~X.}\ \bibnamefont {Fang}}, \bibinfo {author}
  {\bibfnamefont {P.}~\bibnamefont {Avouris}}, \bibinfo {author} {\bibfnamefont
  {T.~F.}\ \bibnamefont {Heinz}}, \bibinfo {author} {\bibfnamefont
  {F.}~\bibnamefont {Guinea}}, \bibinfo {author} {\bibfnamefont
  {L.}~\bibnamefont {Martin-Moreno}}, \ and\ \bibinfo {author} {\bibfnamefont
  {F.}~\bibnamefont {Koppens}},\ }\bibfield  {title} {\bibinfo {title} {\emph
  {Polaritons in layered two-dimensional materials}},\ }\href
  {https://doi.org/10.1038/nmat4792} {\bibfield  {journal} {\bibinfo  {journal}
  {Nature Materials}\ }\textbf {\bibinfo {volume} {16}},\ \bibinfo {pages}
  {182} (\bibinfo {year} {2016})}\BibitemShut {NoStop}%
\bibitem [{\citenamefont {Liu}\ \emph {et~al.}(2014)\citenamefont {Liu},
  \citenamefont {Galfsky}, \citenamefont {Sun}, \citenamefont {Xia},
  \citenamefont {Lin}, \citenamefont {Lee}, \citenamefont {K{\'{e}}na-Cohen},\
  and\ \citenamefont {Menon}}]{Liu2014}%
  \BibitemOpen
  \bibfield  {author} {\bibinfo {author} {\bibfnamefont {X.}~\bibnamefont
  {Liu}}, \bibinfo {author} {\bibfnamefont {T.}~\bibnamefont {Galfsky}},
  \bibinfo {author} {\bibfnamefont {Z.}~\bibnamefont {Sun}}, \bibinfo {author}
  {\bibfnamefont {F.}~\bibnamefont {Xia}}, \bibinfo {author} {\bibfnamefont
  {E.~C.}\ \bibnamefont {Lin}}, \bibinfo {author} {\bibfnamefont {Y.~H.}\
  \bibnamefont {Lee}}, \bibinfo {author} {\bibfnamefont {S.}~\bibnamefont
  {K{\'{e}}na-Cohen}}, \ and\ \bibinfo {author} {\bibfnamefont {V.~M.}\
  \bibnamefont {Menon}},\ }\bibfield  {title} {\bibinfo {title} {\emph {{Strong
  light-matter coupling in two-dimensional atomic crystals}}},\ }\href
  {\doibase 10.1038/nphoton.2014.304} {\bibfield  {journal} {\bibinfo
  {journal} {Nature Photonics}\ }\textbf {\bibinfo {volume} {9}},\ \bibinfo
  {pages} {30} (\bibinfo {year} {2014})}\BibitemShut {NoStop}%
\bibitem [{\citenamefont {Dufferwiel}\ \emph {et~al.}(2015)\citenamefont
  {Dufferwiel}, \citenamefont {Schwarz}, \citenamefont {Withers}, \citenamefont
  {Trichet}, \citenamefont {Li}, \citenamefont {Sich}, \citenamefont {Del
  Pozo-Zamudio}, \citenamefont {Clark}, \citenamefont {Nalitov}, \citenamefont
  {Solnyshkov}, \citenamefont {Malpuech}, \citenamefont {Novoselov},
  \citenamefont {Smith}, \citenamefont {Skolnick}, \citenamefont
  {Krizhanovskii},\ and\ \citenamefont {Tartakovskii}}]{Dufferwiel2015}%
  \BibitemOpen
  \bibfield  {author} {\bibinfo {author} {\bibfnamefont {S.}~\bibnamefont
  {Dufferwiel}}, \bibinfo {author} {\bibfnamefont {S.}~\bibnamefont {Schwarz}},
  \bibinfo {author} {\bibfnamefont {F.}~\bibnamefont {Withers}}, \bibinfo
  {author} {\bibfnamefont {A.~A.~P.}\ \bibnamefont {Trichet}}, \bibinfo
  {author} {\bibfnamefont {F.}~\bibnamefont {Li}}, \bibinfo {author}
  {\bibfnamefont {M.}~\bibnamefont {Sich}}, \bibinfo {author} {\bibfnamefont
  {O.}~\bibnamefont {Del Pozo-Zamudio}}, \bibinfo {author} {\bibfnamefont
  {C.}~\bibnamefont {Clark}}, \bibinfo {author} {\bibfnamefont
  {A.}~\bibnamefont {Nalitov}}, \bibinfo {author} {\bibfnamefont {D.~D.}\
  \bibnamefont {Solnyshkov}}, \bibinfo {author} {\bibfnamefont
  {G.}~\bibnamefont {Malpuech}}, \bibinfo {author} {\bibfnamefont {K.~S.}\
  \bibnamefont {Novoselov}}, \bibinfo {author} {\bibfnamefont {J.~M.}\
  \bibnamefont {Smith}}, \bibinfo {author} {\bibfnamefont {M.~S.}\ \bibnamefont
  {Skolnick}}, \bibinfo {author} {\bibfnamefont {D.~N.}\ \bibnamefont
  {Krizhanovskii}}, \ and\ \bibinfo {author} {\bibfnamefont {A.~I.}\
  \bibnamefont {Tartakovskii}},\ }\bibfield  {title} {\bibinfo {title} {\emph
  {Exciton--polaritons in van der Waals heterostructures embedded in tunable
  microcavities}},\ }\href {https://doi.org/10.1038/ncomms9579} {\bibfield
  {journal} {\bibinfo  {journal} {Nature Communications}\ }\textbf {\bibinfo
  {volume} {6}},\ \bibinfo {pages} {8579} (\bibinfo {year} {2015})}\BibitemShut
  {NoStop}%
\bibitem [{\citenamefont {Sidler}\ \emph {et~al.}(2016)\citenamefont {Sidler},
  \citenamefont {Back}, \citenamefont {Cotlet}, \citenamefont {Srivastava},
  \citenamefont {Fink}, \citenamefont {Kroner}, \citenamefont {Demler},\ and\
  \citenamefont {Imamoglu}}]{Sidler2016}%
  \BibitemOpen
  \bibfield  {author} {\bibinfo {author} {\bibfnamefont {M.}~\bibnamefont
  {Sidler}}, \bibinfo {author} {\bibfnamefont {P.}~\bibnamefont {Back}},
  \bibinfo {author} {\bibfnamefont {O.}~\bibnamefont {Cotlet}}, \bibinfo
  {author} {\bibfnamefont {A.}~\bibnamefont {Srivastava}}, \bibinfo {author}
  {\bibfnamefont {T.}~\bibnamefont {Fink}}, \bibinfo {author} {\bibfnamefont
  {M.}~\bibnamefont {Kroner}}, \bibinfo {author} {\bibfnamefont
  {E.}~\bibnamefont {Demler}}, \ and\ \bibinfo {author} {\bibfnamefont
  {A.}~\bibnamefont {Imamoglu}},\ }\bibfield  {title} {\bibinfo {title} {\emph
  {Fermi polaron-polaritons in charge-tunable atomically thin
  semiconductors}},\ }\href {https://doi.org/10.1038/nphys3949} {\bibfield
  {journal} {\bibinfo  {journal} {Nature Physics}\ }\textbf {\bibinfo {volume}
  {13}},\ \bibinfo {pages} {255} (\bibinfo {year} {2016})}\BibitemShut
  {NoStop}%
\bibitem [{\citenamefont {Kn{\"u}ppel}\ \emph {et~al.}(2019)\citenamefont
  {Kn{\"u}ppel}, \citenamefont {Ravets}, \citenamefont {Kroner}, \citenamefont
  {F{\"a}lt}, \citenamefont {Wegscheider},\ and\ \citenamefont
  {Imamoglu}}]{Knuppel2019}%
  \BibitemOpen
  \bibfield  {author} {\bibinfo {author} {\bibfnamefont {P.}~\bibnamefont
  {Kn{\"u}ppel}}, \bibinfo {author} {\bibfnamefont {S.}~\bibnamefont {Ravets}},
  \bibinfo {author} {\bibfnamefont {M.}~\bibnamefont {Kroner}}, \bibinfo
  {author} {\bibfnamefont {S.}~\bibnamefont {F{\"a}lt}}, \bibinfo {author}
  {\bibfnamefont {W.}~\bibnamefont {Wegscheider}}, \ and\ \bibinfo {author}
  {\bibfnamefont {A.}~\bibnamefont {Imamoglu}},\ }\bibfield  {title} {\bibinfo
  {title} {\emph {Nonlinear optics in the fractional quantum Hall regime}},\
  }\href {\doibase 10.1038/s41586-019-1356-3} {\bibfield  {journal} {\bibinfo
  {journal} {Nature}\ }\textbf {\bibinfo {volume} {572}},\ \bibinfo {pages}
  {91} (\bibinfo {year} {2019})}\BibitemShut {NoStop}%
\bibitem [{\citenamefont {Horikiri}\ \emph {et~al.}(2017)\citenamefont
  {Horikiri}, \citenamefont {Byrnes}, \citenamefont {Kusudo}, \citenamefont
  {Ishida}, \citenamefont {Matsuo}, \citenamefont {Shikano}, \citenamefont
  {L\"offler}, \citenamefont {H\"ofling}, \citenamefont {Forchel},\ and\
  \citenamefont {Yamamoto}}]{Horikiri2017}%
  \BibitemOpen
  \bibfield  {author} {\bibinfo {author} {\bibfnamefont {T.}~\bibnamefont
  {Horikiri}}, \bibinfo {author} {\bibfnamefont {T.}~\bibnamefont {Byrnes}},
  \bibinfo {author} {\bibfnamefont {K.}~\bibnamefont {Kusudo}}, \bibinfo
  {author} {\bibfnamefont {N.}~\bibnamefont {Ishida}}, \bibinfo {author}
  {\bibfnamefont {Y.}~\bibnamefont {Matsuo}}, \bibinfo {author} {\bibfnamefont
  {Y.}~\bibnamefont {Shikano}}, \bibinfo {author} {\bibfnamefont
  {A.}~\bibnamefont {L\"offler}}, \bibinfo {author} {\bibfnamefont
  {S.}~\bibnamefont {H\"ofling}}, \bibinfo {author} {\bibfnamefont
  {A.}~\bibnamefont {Forchel}}, \ and\ \bibinfo {author} {\bibfnamefont
  {Y.}~\bibnamefont {Yamamoto}},\ }\bibfield  {title} {\bibinfo {title} {\emph
  {Highly excited exciton-polariton condensates}},\ }\href {\doibase
  10.1103/PhysRevB.95.245122} {\bibfield  {journal} {\bibinfo  {journal} {Phys.
  Rev. B}\ }\textbf {\bibinfo {volume} {95}},\ \bibinfo {pages} {245122}
  (\bibinfo {year} {2017})}\BibitemShut {NoStop}%
\bibitem [{\citenamefont {Kamide}\ and\ \citenamefont
  {Ogawa}(2010)}]{Kamide2010}%
  \BibitemOpen
  \bibfield  {author} {\bibinfo {author} {\bibfnamefont {K.}~\bibnamefont
  {Kamide}}\ and\ \bibinfo {author} {\bibfnamefont {T.}~\bibnamefont {Ogawa}},\
  }\bibfield  {title} {\bibinfo {title} {\emph {What Determines the Wave
  Function of Electron-Hole Pairs in Polariton Condensates?}},\ }\href
  {\doibase 10.1103/PhysRevLett.105.056401} {\bibfield  {journal} {\bibinfo
  {journal} {Phys. Rev. Lett.}\ }\textbf {\bibinfo {volume} {105}},\ \bibinfo
  {pages} {056401} (\bibinfo {year} {2010})}\BibitemShut {NoStop}%
\bibitem [{\citenamefont {Byrnes}\ \emph {et~al.}(2010)\citenamefont {Byrnes},
  \citenamefont {Horikiri}, \citenamefont {Ishida},\ and\ \citenamefont
  {Yamamoto}}]{Byrnes2010}%
  \BibitemOpen
  \bibfield  {author} {\bibinfo {author} {\bibfnamefont {T.}~\bibnamefont
  {Byrnes}}, \bibinfo {author} {\bibfnamefont {T.}~\bibnamefont {Horikiri}},
  \bibinfo {author} {\bibfnamefont {N.}~\bibnamefont {Ishida}}, \ and\ \bibinfo
  {author} {\bibfnamefont {Y.}~\bibnamefont {Yamamoto}},\ }\bibfield  {title}
  {\bibinfo {title} {\emph {BCS Wave-Function Approach to the BEC-BCS Crossover
  of Exciton-Polariton Condensates}},\ }\href
  {https://link.aps.org/doi/10.1103/PhysRevLett.105.186402} {\bibfield
  {journal} {\bibinfo  {journal} {Phys. Rev. Lett.}\ }\textbf {\bibinfo
  {volume} {105}},\ \bibinfo {pages} {186402} (\bibinfo {year}
  {2010})}\BibitemShut {NoStop}%
\bibitem [{\citenamefont {Kamide}\ and\ \citenamefont
  {Ogawa}(2011)}]{Kamide2011}%
  \BibitemOpen
  \bibfield  {author} {\bibinfo {author} {\bibfnamefont {K.}~\bibnamefont
  {Kamide}}\ and\ \bibinfo {author} {\bibfnamefont {T.}~\bibnamefont {Ogawa}},\
  }\bibfield  {title} {\bibinfo {title} {\emph {Ground-state properties of
  microcavity polariton condensates at arbitrary excitation density}},\ }\href
  {\doibase 10.1103/PhysRevB.83.165319} {\bibfield  {journal} {\bibinfo
  {journal} {Phys. Rev. B}\ }\textbf {\bibinfo {volume} {83}},\ \bibinfo
  {pages} {165319} (\bibinfo {year} {2011})}\BibitemShut {NoStop}%
\bibitem [{\citenamefont {Yamaguchi}\ \emph {et~al.}(2012)\citenamefont
  {Yamaguchi}, \citenamefont {Kamide}, \citenamefont {Ogawa},\ and\
  \citenamefont {Yamamoto}}]{Yamaguchi2012}%
  \BibitemOpen
  \bibfield  {author} {\bibinfo {author} {\bibfnamefont {M.}~\bibnamefont
  {Yamaguchi}}, \bibinfo {author} {\bibfnamefont {K.}~\bibnamefont {Kamide}},
  \bibinfo {author} {\bibfnamefont {T.}~\bibnamefont {Ogawa}}, \ and\ \bibinfo
  {author} {\bibfnamefont {Y.}~\bibnamefont {Yamamoto}},\ }\bibfield  {title}
  {\bibinfo {title} {\emph {BEC--BCS-laser crossover in Coulomb-correlated
  electron--hole--photon systems}},\ }\href
  {http://stacks.iop.org/1367-2630/14/i=6/a=065001} {\bibfield  {journal}
  {\bibinfo  {journal} {New Journal of Physics}\ }\textbf {\bibinfo {volume}
  {14}},\ \bibinfo {pages} {065001} (\bibinfo {year} {2012})}\BibitemShut
  {NoStop}%
\bibitem [{\citenamefont {Yamaguchi}\ \emph {et~al.}(2013)\citenamefont
  {Yamaguchi}, \citenamefont {Kamide}, \citenamefont {Nii}, \citenamefont
  {Ogawa},\ and\ \citenamefont {Yamamoto}}]{Yamaguchi2013}%
  \BibitemOpen
  \bibfield  {author} {\bibinfo {author} {\bibfnamefont {M.}~\bibnamefont
  {Yamaguchi}}, \bibinfo {author} {\bibfnamefont {K.}~\bibnamefont {Kamide}},
  \bibinfo {author} {\bibfnamefont {R.}~\bibnamefont {Nii}}, \bibinfo {author}
  {\bibfnamefont {T.}~\bibnamefont {Ogawa}}, \ and\ \bibinfo {author}
  {\bibfnamefont {Y.}~\bibnamefont {Yamamoto}},\ }\bibfield  {title} {\bibinfo
  {title} {\emph {Second Thresholds in BEC-BCS-Laser Crossover of
  Exciton-Polariton Systems}},\ }\href {\doibase
  10.1103/PhysRevLett.111.026404} {\bibfield  {journal} {\bibinfo  {journal}
  {Phys. Rev. Lett.}\ }\textbf {\bibinfo {volume} {111}},\ \bibinfo {pages}
  {026404} (\bibinfo {year} {2013})}\BibitemShut {NoStop}%
\bibitem [{\citenamefont {Zhang}\ \emph {et~al.}(2013)\citenamefont {Zhang},
  \citenamefont {Kim}, \citenamefont {Yamamoto},\ and\ \citenamefont
  {Na}}]{Zhang2013}%
  \BibitemOpen
  \bibfield  {author} {\bibinfo {author} {\bibfnamefont {H.}~\bibnamefont
  {Zhang}}, \bibinfo {author} {\bibfnamefont {N.~Y.}\ \bibnamefont {Kim}},
  \bibinfo {author} {\bibfnamefont {Y.}~\bibnamefont {Yamamoto}}, \ and\
  \bibinfo {author} {\bibfnamefont {N.}~\bibnamefont {Na}},\ }\bibfield
  {title} {\bibinfo {title} {\emph {Very strong coupling in GaAs-based optical
  microcavities}},\ }\href {\doibase 10.1103/PhysRevB.87.115303} {\bibfield
  {journal} {\bibinfo  {journal} {Phys. Rev. B}\ }\textbf {\bibinfo {volume}
  {87}},\ \bibinfo {pages} {115303} (\bibinfo {year} {2013})}\BibitemShut
  {NoStop}%
\bibitem [{\citenamefont {Yang}\ \emph {et~al.}(2015)\citenamefont {Yang},
  \citenamefont {Kim}, \citenamefont {Yamamoto},\ and\ \citenamefont
  {Na}}]{Yang2015}%
  \BibitemOpen
  \bibfield  {author} {\bibinfo {author} {\bibfnamefont {M.-J.}\ \bibnamefont
  {Yang}}, \bibinfo {author} {\bibfnamefont {N.~Y.}\ \bibnamefont {Kim}},
  \bibinfo {author} {\bibfnamefont {Y.}~\bibnamefont {Yamamoto}}, \ and\
  \bibinfo {author} {\bibfnamefont {N.}~\bibnamefont {Na}},\ }\bibfield
  {title} {\bibinfo {title} {\emph {Verification of very strong coupling in a
  semiconductor optical microcavity}},\ }\href
  {http://stacks.iop.org/1367-2630/17/i=2/a=023064} {\bibfield  {journal}
  {\bibinfo  {journal} {New Journal of Physics}\ }\textbf {\bibinfo {volume}
  {17}},\ \bibinfo {pages} {023064} (\bibinfo {year} {2015})}\BibitemShut
  {NoStop}%
\bibitem [{\citenamefont {Averkiev}\ and\ \citenamefont
  {Glazov}(2007)}]{Averkiev2007}%
  \BibitemOpen
  \bibfield  {author} {\bibinfo {author} {\bibfnamefont {N.~S.}\ \bibnamefont
  {Averkiev}}\ and\ \bibinfo {author} {\bibfnamefont {M.~M.}\ \bibnamefont
  {Glazov}},\ }\bibfield  {title} {\bibinfo {title} {\emph {Light-matter
  interaction in doped microcavities}},\ }\href {\doibase
  10.1103/PhysRevB.76.045320} {\bibfield  {journal} {\bibinfo  {journal} {Phys.
  Rev. B}\ }\textbf {\bibinfo {volume} {76}},\ \bibinfo {pages} {045320}
  (\bibinfo {year} {2007})}\BibitemShut {NoStop}%
\bibitem [{\citenamefont {Brodbeck}\ \emph {et~al.}(2017)\citenamefont
  {Brodbeck}, \citenamefont {De~Liberato}, \citenamefont {Amthor},
  \citenamefont {Klaas}, \citenamefont {Kamp}, \citenamefont {Worschech},
  \citenamefont {Schneider},\ and\ \citenamefont {H\"ofling}}]{Brodbeck2017}%
  \BibitemOpen
  \bibfield  {author} {\bibinfo {author} {\bibfnamefont {S.}~\bibnamefont
  {Brodbeck}}, \bibinfo {author} {\bibfnamefont {S.}~\bibnamefont
  {De~Liberato}}, \bibinfo {author} {\bibfnamefont {M.}~\bibnamefont {Amthor}},
  \bibinfo {author} {\bibfnamefont {M.}~\bibnamefont {Klaas}}, \bibinfo
  {author} {\bibfnamefont {M.}~\bibnamefont {Kamp}}, \bibinfo {author}
  {\bibfnamefont {L.}~\bibnamefont {Worschech}}, \bibinfo {author}
  {\bibfnamefont {C.}~\bibnamefont {Schneider}}, \ and\ \bibinfo {author}
  {\bibfnamefont {S.}~\bibnamefont {H\"ofling}},\ }\bibfield  {title} {\bibinfo
  {title} {\emph {Experimental Verification of the Very Strong Coupling Regime
  in a GaAs Quantum Well Microcavity}},\ }\href {\doibase
  10.1103/PhysRevLett.119.027401} {\bibfield  {journal} {\bibinfo  {journal}
  {Phys. Rev. Lett.}\ }\textbf {\bibinfo {volume} {119}},\ \bibinfo {pages}
  {027401} (\bibinfo {year} {2017})}\BibitemShut {NoStop}%
\bibitem [{\citenamefont {{Combescot}}\ \emph {et~al.}(2008)\citenamefont
  {{Combescot}}, \citenamefont {{Betbeder-Matibet}},\ and\ \citenamefont
  {{Dubin}}}]{MCombescot2008}%
  \BibitemOpen
  \bibfield  {author} {\bibinfo {author} {\bibfnamefont {M.}~\bibnamefont
  {{Combescot}}}, \bibinfo {author} {\bibfnamefont {O.}~\bibnamefont
  {{Betbeder-Matibet}}}, \ and\ \bibinfo {author} {\bibfnamefont
  {F.}~\bibnamefont {{Dubin}}},\ }\bibfield  {title} {\bibinfo {title} {\emph
  {{The many-body physics of composite bosons}}},\ }\href {\doibase
  10.1016/j.physrep.2007.11.003} {\bibfield  {journal} {\bibinfo  {journal}
  {Physics Reports}\ }\textbf {\bibinfo {volume} {463}},\ \bibinfo {pages}
  {215} (\bibinfo {year} {2008})}\BibitemShut {NoStop}%
\bibitem [{\citenamefont {Glazov}\ \emph {et~al.}(2009)\citenamefont {Glazov},
  \citenamefont {Ouerdane}, \citenamefont {Pilozzi}, \citenamefont {Malpuech},
  \citenamefont {Kavokin},\ and\ \citenamefont {D'Andrea}}]{Glazov2009}%
  \BibitemOpen
  \bibfield  {author} {\bibinfo {author} {\bibfnamefont {M.~M.}\ \bibnamefont
  {Glazov}}, \bibinfo {author} {\bibfnamefont {H.}~\bibnamefont {Ouerdane}},
  \bibinfo {author} {\bibfnamefont {L.}~\bibnamefont {Pilozzi}}, \bibinfo
  {author} {\bibfnamefont {G.}~\bibnamefont {Malpuech}}, \bibinfo {author}
  {\bibfnamefont {A.~V.}\ \bibnamefont {Kavokin}}, \ and\ \bibinfo {author}
  {\bibfnamefont {A.}~\bibnamefont {D'Andrea}},\ }\bibfield  {title} {\bibinfo
  {title} {\emph {Polariton-polariton scattering in microcavities: A
  microscopic theory}},\ }\href {\doibase 10.1103/PhysRevB.80.155306}
  {\bibfield  {journal} {\bibinfo  {journal} {Phys. Rev. B}\ }\textbf {\bibinfo
  {volume} {80}},\ \bibinfo {pages} {155306} (\bibinfo {year}
  {2009})}\BibitemShut {NoStop}%
\bibitem [{\citenamefont {Parfitt}\ and\ \citenamefont
  {Portnoi}(2002)}]{2DexcitonEnergy}%
  \BibitemOpen
  \bibfield  {author} {\bibinfo {author} {\bibfnamefont {D.~G.~W.}\
  \bibnamefont {Parfitt}}\ and\ \bibinfo {author} {\bibfnamefont {M.~E.}\
  \bibnamefont {Portnoi}},\ }\bibfield  {title} {\bibinfo {title} {\emph {The
  two-dimensional hydrogen atom revisited}},\ }\href {\doibase
  10.1063/1.1503868} {\bibfield  {journal} {\bibinfo  {journal} {Journal of
  Mathematical Physics}\ }\textbf {\bibinfo {volume} {43}},\ \bibinfo {pages}
  {4681} (\bibinfo {year} {2002})}\BibitemShut {NoStop}%
\bibitem [{\citenamefont {Hopfield}(1958)}]{Hopfield1958}%
  \BibitemOpen
  \bibfield  {author} {\bibinfo {author} {\bibfnamefont {J.~J.}\ \bibnamefont
  {Hopfield}},\ }\bibfield  {title} {\bibinfo {title} {\emph {Theory of the
  Contribution of Excitons to the Complex Dielectric Constant of Crystals}},\
  }\href {\doibase 10.1103/PhysRev.112.1555} {\bibfield  {journal} {\bibinfo
  {journal} {Phys. Rev.}\ }\textbf {\bibinfo {volume} {112}},\ \bibinfo {pages}
  {1555} (\bibinfo {year} {1958})}\BibitemShut {NoStop}%
\bibitem [{\citenamefont {Weinberg}(2005)}]{WeinbergVol1}%
  \BibitemOpen
  \bibfield  {author} {\bibinfo {author} {\bibfnamefont {S.}~\bibnamefont
  {Weinberg}},\ }\href@noop {} {\emph {\bibinfo {title} {{The Quantum theory of
  fields. Vol. 1: Foundations}}}}\ (\bibinfo  {publisher} {Cambridge University
  Press, Cambridge UK},\ \bibinfo {year} {2005})\BibitemShut {NoStop}%
\bibitem [{\citenamefont {Boyd}(1999)}]{Boyd1999}%
  \BibitemOpen
  \bibfield  {author} {\bibinfo {author} {\bibfnamefont {J.~P.}\ \bibnamefont
  {Boyd}},\ }\bibfield  {title} {\bibinfo {title} {\emph {The Devil's
  Invention: Asymptotic, Superasymptotic and Hyperasymptotic Series}},\ }\href
  {\doibase 10.1023/A:1006145903624} {\bibfield  {journal} {\bibinfo  {journal}
  {Acta Applicandae Mathematica}\ }\textbf {\bibinfo {volume} {56}},\ \bibinfo
  {pages} {1} (\bibinfo {year} {1999})}\BibitemShut {NoStop}%
\bibitem [{\citenamefont {Fetter}\ and\ \citenamefont
  {Walecka}(2003)}]{fetterbook}%
  \BibitemOpen
  \bibfield  {author} {\bibinfo {author} {\bibfnamefont {A.~L.}\ \bibnamefont
  {Fetter}}\ and\ \bibinfo {author} {\bibfnamefont {J.~D.}\ \bibnamefont
  {Walecka}},\ }\href@noop {} {\emph {\bibinfo {title} {Quantum theory of
  many-particle systems}}}\ (\bibinfo  {publisher} {Dover Publications,
  Mineola, New York USA},\ \bibinfo {year} {2003})\BibitemShut {NoStop}%
\bibitem [{\citenamefont {Dittrich}(1999)}]{2DGreenfunction}%
  \BibitemOpen
  \bibfield  {author} {\bibinfo {author} {\bibfnamefont {W.}~\bibnamefont
  {Dittrich}},\ }\bibfield  {title} {\bibinfo {title} {\emph {The Coulomb
  Green's function in two dimensions}},\ }\href {\doibase 10.1119/1.19123}
  {\bibfield  {journal} {\bibinfo  {journal} {American Journal of Physics}\
  }\textbf {\bibinfo {volume} {67}},\ \bibinfo {pages} {768} (\bibinfo {year}
  {1999})}\BibitemShut {NoStop}%
\bibitem [{\citenamefont {Press}\ \emph {et~al.}(2007)\citenamefont {Press},
  \citenamefont {Teukolsky}, \citenamefont {Vetterling},\ and\ \citenamefont
  {Flannery}}]{numericalrecipes}%
  \BibitemOpen
  \bibfield  {author} {\bibinfo {author} {\bibfnamefont {W.}~\bibnamefont
  {Press}}, \bibinfo {author} {\bibfnamefont {S.}~\bibnamefont {Teukolsky}},
  \bibinfo {author} {\bibfnamefont {W.}~\bibnamefont {Vetterling}}, \ and\
  \bibinfo {author} {\bibfnamefont {B.}~\bibnamefont {Flannery}},\ }\bibfield
  {title} {\bibinfo {title} {\emph {Numerical recipes: The art of scientific
  computing}},\ }\href@noop {} {\bibfield  {journal} {\bibinfo  {journal}
  {Cambridge University Press, Cambridge UK}\ } (\bibinfo {year}
  {2007})}\BibitemShut {NoStop}%
\bibitem [{\citenamefont {Sze}(2007)}]{EffectiveMass}%
  \BibitemOpen
  \bibfield  {author} {\bibinfo {author} {\bibfnamefont {S.~M.}\ \bibnamefont
  {Sze}},\ }\href@noop {} {\emph {\bibinfo {title} {Physics of semiconductor
  devices}}},\ \bibinfo {edition} {3rd}\ ed.\ (\bibinfo  {publisher} {Hoboken,
  NJ : Wiley-Interscience},\ \bibinfo {address} {Hoboken, NJ Hoboken, N.J.},\
  \bibinfo {year} {2007})\BibitemShut {NoStop}%
\bibitem [{\citenamefont {Sanvitto}\ and\ \citenamefont
  {K{\'e}na-Cohen}(2016)}]{Sanvitto2016}%
  \BibitemOpen
  \bibfield  {author} {\bibinfo {author} {\bibfnamefont {D.}~\bibnamefont
  {Sanvitto}}\ and\ \bibinfo {author} {\bibfnamefont {S.}~\bibnamefont
  {K{\'e}na-Cohen}},\ }\bibfield  {title} {\bibinfo {title} {\emph {The road
  towards polaritonic devices}},\ }\href {https://doi.org/10.1038/nmat4668}
  {\bibfield  {journal} {\bibinfo  {journal} {Nature Materials}\ }\textbf
  {\bibinfo {volume} {15}},\ \bibinfo {pages} {1061} (\bibinfo {year}
  {2016})}\BibitemShut {NoStop}%
\bibitem [{\citenamefont {Guillet}\ and\ \citenamefont
  {Brimont}(2016)}]{Guillet2016}%
  \BibitemOpen
  \bibfield  {author} {\bibinfo {author} {\bibfnamefont {T.}~\bibnamefont
  {Guillet}}\ and\ \bibinfo {author} {\bibfnamefont {C.}~\bibnamefont
  {Brimont}},\ }\bibfield  {title} {\bibinfo {title} {\emph {Polariton
  condensates at room temperature}},\ }\href {\doibase
  10.1016/j.crhy.2016.07.002} {\bibfield  {journal} {\bibinfo  {journal}
  {Comptes Rendus Physique}\ }\textbf {\bibinfo {volume} {17}},\ \bibinfo
  {pages} {946 } (\bibinfo {year} {2016})}\BibitemShut {NoStop}%
\bibitem [{\citenamefont {Ciuti}\ and\ \citenamefont
  {Carusotto}(2006)}]{CiutiPRA2006}%
  \BibitemOpen
  \bibfield  {author} {\bibinfo {author} {\bibfnamefont {C.}~\bibnamefont
  {Ciuti}}\ and\ \bibinfo {author} {\bibfnamefont {I.}~\bibnamefont
  {Carusotto}},\ }\bibfield  {title} {\bibinfo {title} {\emph {Input-output
  theory of cavities in the ultrastrong coupling regime: The case of
  time-independent cavity parameters}},\ }\href {\doibase
  10.1103/PhysRevA.74.033811} {\bibfield  {journal} {\bibinfo  {journal} {Phys.
  Rev. A}\ }\textbf {\bibinfo {volume} {74}},\ \bibinfo {pages} {033811}
  (\bibinfo {year} {2006})}\BibitemShut {NoStop}%
\bibitem [{\citenamefont {\ifmmode~\acute{C}\else \'{C}\fi{}wik}\ \emph
  {et~al.}(2016)\citenamefont {\ifmmode~\acute{C}\else \'{C}\fi{}wik},
  \citenamefont {Kirton}, \citenamefont {De~Liberato},\ and\ \citenamefont
  {Keeling}}]{Cwik2016}%
  \BibitemOpen
  \bibfield  {author} {\bibinfo {author} {\bibfnamefont {J.~A.}\ \bibnamefont
  {\ifmmode~\acute{C}\else \'{C}\fi{}wik}}, \bibinfo {author} {\bibfnamefont
  {P.}~\bibnamefont {Kirton}}, \bibinfo {author} {\bibfnamefont
  {S.}~\bibnamefont {De~Liberato}}, \ and\ \bibinfo {author} {\bibfnamefont
  {J.}~\bibnamefont {Keeling}},\ }\bibfield  {title} {\bibinfo {title} {\emph
  {Excitonic spectral features in strongly coupled organic polaritons}},\
  }\href {\doibase 10.1103/PhysRevA.93.033840} {\bibfield  {journal} {\bibinfo
  {journal} {Phys. Rev. A}\ }\textbf {\bibinfo {volume} {93}},\ \bibinfo
  {pages} {033840} (\bibinfo {year} {2016})}\BibitemShut {NoStop}%
\bibitem [{\citenamefont {Citrin}\ and\ \citenamefont
  {Khurgin}(2003)}]{Citrin2003}%
  \BibitemOpen
  \bibfield  {author} {\bibinfo {author} {\bibfnamefont {D.~S.}\ \bibnamefont
  {Citrin}}\ and\ \bibinfo {author} {\bibfnamefont {J.~B.}\ \bibnamefont
  {Khurgin}},\ }\bibfield  {title} {\bibinfo {title} {\emph {Microcavity effect
  on the electron-hole relative motion in semiconductor quantum wells}},\
  }\href {\doibase 10.1103/PhysRevB.68.205325} {\bibfield  {journal} {\bibinfo
  {journal} {Phys. Rev. B}\ }\textbf {\bibinfo {volume} {68}},\ \bibinfo
  {pages} {205325} (\bibinfo {year} {2003})}\BibitemShut {NoStop}%
\bibitem [{\citenamefont {Kazimierczuk}\ \emph {et~al.}(2014)\citenamefont
  {Kazimierczuk}, \citenamefont {Fr{\"o}hlich}, \citenamefont {Scheel},
  \citenamefont {Stolz},\ and\ \citenamefont {Bayer}}]{Kazimierczuk2014}%
  \BibitemOpen
  \bibfield  {author} {\bibinfo {author} {\bibfnamefont {T.}~\bibnamefont
  {Kazimierczuk}}, \bibinfo {author} {\bibfnamefont {D.}~\bibnamefont
  {Fr{\"o}hlich}}, \bibinfo {author} {\bibfnamefont {S.}~\bibnamefont
  {Scheel}}, \bibinfo {author} {\bibfnamefont {H.}~\bibnamefont {Stolz}}, \
  and\ \bibinfo {author} {\bibfnamefont {M.}~\bibnamefont {Bayer}},\ }\bibfield
   {title} {\bibinfo {title} {\emph {Giant Rydberg excitons in the copper oxide
  Cu$_2$O}},\ }\href {https://doi.org/10.1038/nature13832} {\bibfield
  {journal} {\bibinfo  {journal} {Nature}\ }\textbf {\bibinfo {volume} {514}},\
  \bibinfo {pages} {343} (\bibinfo {year} {2014})}\BibitemShut {NoStop}%
\bibitem [{\citenamefont {Ferrier}\ \emph {et~al.}(2011)\citenamefont
  {Ferrier}, \citenamefont {Wertz}, \citenamefont {Johne}, \citenamefont
  {Solnyshkov}, \citenamefont {Senellart}, \citenamefont {Sagnes},
  \citenamefont {Lema\^{\i}tre}, \citenamefont {Malpuech},\ and\ \citenamefont
  {Bloch}}]{FerrierPRL2011}%
  \BibitemOpen
  \bibfield  {author} {\bibinfo {author} {\bibfnamefont {L.}~\bibnamefont
  {Ferrier}}, \bibinfo {author} {\bibfnamefont {E.}~\bibnamefont {Wertz}},
  \bibinfo {author} {\bibfnamefont {R.}~\bibnamefont {Johne}}, \bibinfo
  {author} {\bibfnamefont {D.~D.}\ \bibnamefont {Solnyshkov}}, \bibinfo
  {author} {\bibfnamefont {P.}~\bibnamefont {Senellart}}, \bibinfo {author}
  {\bibfnamefont {I.}~\bibnamefont {Sagnes}}, \bibinfo {author} {\bibfnamefont
  {A.}~\bibnamefont {Lema\^{\i}tre}}, \bibinfo {author} {\bibfnamefont
  {G.}~\bibnamefont {Malpuech}}, \ and\ \bibinfo {author} {\bibfnamefont
  {J.}~\bibnamefont {Bloch}},\ }\bibfield  {title} {\bibinfo {title} {\emph
  {Interactions in Confined Polariton Condensates}},\ }\href {\doibase
  10.1103/PhysRevLett.106.126401} {\bibfield  {journal} {\bibinfo  {journal}
  {Phys. Rev. Lett.}\ }\textbf {\bibinfo {volume} {106}},\ \bibinfo {pages}
  {126401} (\bibinfo {year} {2011})}\BibitemShut {NoStop}%
\bibitem [{\citenamefont {Brichkin}\ \emph {et~al.}(2011)\citenamefont
  {Brichkin}, \citenamefont {Novikov}, \citenamefont {Larionov}, \citenamefont
  {Kulakovskii}, \citenamefont {Glazov}, \citenamefont {Schneider},
  \citenamefont {H{\"o}fling}, \citenamefont {Kamp},\ and\ \citenamefont
  {Forchel}}]{Brichkin2011}%
  \BibitemOpen
  \bibfield  {author} {\bibinfo {author} {\bibfnamefont {A.~S.}\ \bibnamefont
  {Brichkin}}, \bibinfo {author} {\bibfnamefont {S.~I.}\ \bibnamefont
  {Novikov}}, \bibinfo {author} {\bibfnamefont {A.~V.}\ \bibnamefont
  {Larionov}}, \bibinfo {author} {\bibfnamefont {V.~D.}\ \bibnamefont
  {Kulakovskii}}, \bibinfo {author} {\bibfnamefont {M.~M.}\ \bibnamefont
  {Glazov}}, \bibinfo {author} {\bibfnamefont {C.}~\bibnamefont {Schneider}},
  \bibinfo {author} {\bibfnamefont {S.}~\bibnamefont {H{\"o}fling}}, \bibinfo
  {author} {\bibfnamefont {M.}~\bibnamefont {Kamp}}, \ and\ \bibinfo {author}
  {\bibfnamefont {A.}~\bibnamefont {Forchel}},\ }\bibfield  {title} {\bibinfo
  {title} {\emph {Effect of Coulomb interaction on exciton-polariton
  condensates in GaAs pillar microcavities}},\ }\href {\doibase
  10.1103/PhysRevB.84.195301} {\bibfield  {journal} {\bibinfo  {journal} {Phys.
  Rev. B}\ }\textbf {\bibinfo {volume} {84}},\ \bibinfo {pages} {195301}
  (\bibinfo {year} {2011})}\BibitemShut {NoStop}%
\bibitem [{\citenamefont {Rodriguez}\ \emph {et~al.}(2016)\citenamefont
  {Rodriguez}, \citenamefont {Amo}, \citenamefont {Sagnes}, \citenamefont
  {Le~Gratiet}, \citenamefont {Galopin}, \citenamefont {Lema{\^\i}tre},\ and\
  \citenamefont {Bloch}}]{Rodriguez2016}%
  \BibitemOpen
  \bibfield  {author} {\bibinfo {author} {\bibfnamefont {S.~R.~K.}\
  \bibnamefont {Rodriguez}}, \bibinfo {author} {\bibfnamefont {A.}~\bibnamefont
  {Amo}}, \bibinfo {author} {\bibfnamefont {I.}~\bibnamefont {Sagnes}},
  \bibinfo {author} {\bibfnamefont {L.}~\bibnamefont {Le~Gratiet}}, \bibinfo
  {author} {\bibfnamefont {E.}~\bibnamefont {Galopin}}, \bibinfo {author}
  {\bibfnamefont {A.}~\bibnamefont {Lema{\^\i}tre}}, \ and\ \bibinfo {author}
  {\bibfnamefont {J.}~\bibnamefont {Bloch}},\ }\bibfield  {title} {\bibinfo
  {title} {\emph {Interaction-induced hopping phase in driven-dissipative
  coupled photonic microcavities}},\ }\href
  {https://doi.org/10.1038/ncomms11887} {\bibfield  {journal} {\bibinfo
  {journal} {Nature Communications}\ }\textbf {\bibinfo {volume} {7}},\
  \bibinfo {pages} {11887} (\bibinfo {year} {2016})}\BibitemShut {NoStop}%
\bibitem [{\citenamefont {Walker}\ \emph {et~al.}(2017)\citenamefont {Walker},
  \citenamefont {Tinkler}, \citenamefont {Royall}, \citenamefont {Skryabin},
  \citenamefont {Farrer}, \citenamefont {Ritchie}, \citenamefont {Skolnick},\
  and\ \citenamefont {Krizhanovskii}}]{WalkerPRL2017}%
  \BibitemOpen
  \bibfield  {author} {\bibinfo {author} {\bibfnamefont {P.~M.}\ \bibnamefont
  {Walker}}, \bibinfo {author} {\bibfnamefont {L.}~\bibnamefont {Tinkler}},
  \bibinfo {author} {\bibfnamefont {B.}~\bibnamefont {Royall}}, \bibinfo
  {author} {\bibfnamefont {D.~V.}\ \bibnamefont {Skryabin}}, \bibinfo {author}
  {\bibfnamefont {I.}~\bibnamefont {Farrer}}, \bibinfo {author} {\bibfnamefont
  {D.~A.}\ \bibnamefont {Ritchie}}, \bibinfo {author} {\bibfnamefont {M.~S.}\
  \bibnamefont {Skolnick}}, \ and\ \bibinfo {author} {\bibfnamefont {D.~N.}\
  \bibnamefont {Krizhanovskii}},\ }\bibfield  {title} {\bibinfo {title} {\emph
  {Dark Solitons in High Velocity Waveguide Polariton Fluids}},\ }\href
  {\doibase 10.1103/PhysRevLett.119.097403} {\bibfield  {journal} {\bibinfo
  {journal} {Phys. Rev. Lett.}\ }\textbf {\bibinfo {volume} {119}},\ \bibinfo
  {pages} {097403} (\bibinfo {year} {2017})}\BibitemShut {NoStop}%
\bibitem [{\citenamefont {Estrecho}\ \emph {et~al.}(2019)\citenamefont
  {Estrecho}, \citenamefont {Gao}, \citenamefont {Bobrovska}, \citenamefont
  {Comber-Todd}, \citenamefont {Fraser}, \citenamefont {Steger}, \citenamefont
  {West}, \citenamefont {Pfeiffer}, \citenamefont {Levinsen}, \citenamefont
  {Parish}, \citenamefont {Liew}, \citenamefont {Matuszewski}, \citenamefont
  {Snoke}, \citenamefont {Truscott},\ and\ \citenamefont
  {Ostrovskaya}}]{Estrecho2018}%
  \BibitemOpen
  \bibfield  {author} {\bibinfo {author} {\bibfnamefont {E.}~\bibnamefont
  {Estrecho}}, \bibinfo {author} {\bibfnamefont {T.}~\bibnamefont {Gao}},
  \bibinfo {author} {\bibfnamefont {N.}~\bibnamefont {Bobrovska}}, \bibinfo
  {author} {\bibfnamefont {D.}~\bibnamefont {Comber-Todd}}, \bibinfo {author}
  {\bibfnamefont {M.~D.}\ \bibnamefont {Fraser}}, \bibinfo {author}
  {\bibfnamefont {M.}~\bibnamefont {Steger}}, \bibinfo {author} {\bibfnamefont
  {K.}~\bibnamefont {West}}, \bibinfo {author} {\bibfnamefont {L.~N.}\
  \bibnamefont {Pfeiffer}}, \bibinfo {author} {\bibfnamefont {J.}~\bibnamefont
  {Levinsen}}, \bibinfo {author} {\bibfnamefont {M.~M.}\ \bibnamefont
  {Parish}}, \bibinfo {author} {\bibfnamefont {T.~C.~H.}\ \bibnamefont {Liew}},
  \bibinfo {author} {\bibfnamefont {M.}~\bibnamefont {Matuszewski}}, \bibinfo
  {author} {\bibfnamefont {D.~W.}\ \bibnamefont {Snoke}}, \bibinfo {author}
  {\bibfnamefont {A.~G.}\ \bibnamefont {Truscott}}, \ and\ \bibinfo {author}
  {\bibfnamefont {E.~A.}\ \bibnamefont {Ostrovskaya}},\ }\bibfield  {title}
  {\bibinfo {title} {\emph {Direct measurement of polariton-polariton
  interaction strength in the Thomas-Fermi regime of exciton-polariton
  condensation}},\ }\href {\doibase 10.1103/PhysRevB.100.035306} {\bibfield
  {journal} {\bibinfo  {journal} {Phys. Rev. B}\ }\textbf {\bibinfo {volume}
  {100}},\ \bibinfo {pages} {035306} (\bibinfo {year} {2019})}\BibitemShut
  {NoStop}%
\bibitem [{\citenamefont {Wang}\ \emph {et~al.}(2018)\citenamefont {Wang},
  \citenamefont {Chernikov}, \citenamefont {Glazov}, \citenamefont {Heinz},
  \citenamefont {Marie}, \citenamefont {Amand},\ and\ \citenamefont
  {Urbaszek}}]{Wang2018}%
  \BibitemOpen
  \bibfield  {author} {\bibinfo {author} {\bibfnamefont {G.}~\bibnamefont
  {Wang}}, \bibinfo {author} {\bibfnamefont {A.}~\bibnamefont {Chernikov}},
  \bibinfo {author} {\bibfnamefont {M.~M.}\ \bibnamefont {Glazov}}, \bibinfo
  {author} {\bibfnamefont {T.~F.}\ \bibnamefont {Heinz}}, \bibinfo {author}
  {\bibfnamefont {X.}~\bibnamefont {Marie}}, \bibinfo {author} {\bibfnamefont
  {T.}~\bibnamefont {Amand}}, \ and\ \bibinfo {author} {\bibfnamefont
  {B.}~\bibnamefont {Urbaszek}},\ }\bibfield  {title} {\bibinfo {title} {\emph
  {Colloquium: Excitons in atomically thin transition metal dichalcogenides}},\
  }\href {\doibase 10.1103/RevModPhys.90.021001} {\bibfield  {journal}
  {\bibinfo  {journal} {Rev. Mod. Phys.}\ }\textbf {\bibinfo {volume} {90}},\
  \bibinfo {pages} {021001} (\bibinfo {year} {2018})}\BibitemShut {NoStop}%
\bibitem [{\citenamefont {{Adhikari}}(1986)}]{Adhikari1986}%
  \BibitemOpen
  \bibfield  {author} {\bibinfo {author} {\bibfnamefont {S.~K.}\ \bibnamefont
  {{Adhikari}}},\ }\bibfield  {title} {\bibinfo {title} {\emph {{Quantum
  scattering in two dimensions}}},\ }\href {\doibase 10.1119/1.14623}
  {\bibfield  {journal} {\bibinfo  {journal} {American Journal of Physics}\
  }\textbf {\bibinfo {volume} {54}},\ \bibinfo {pages} {362} (\bibinfo {year}
  {1986})}\BibitemShut {NoStop}%
\bibitem [{\citenamefont {Levinsen}\ and\ \citenamefont {Parish}()}]{2Dreview}%
  \BibitemOpen
  \bibfield  {author} {\bibinfo {author} {\bibfnamefont {J.}~\bibnamefont
  {Levinsen}}\ and\ \bibinfo {author} {\bibfnamefont {M.~M.}\ \bibnamefont
  {Parish}},\ }\bibinfo {title} {\emph {Strongly interacting two-dimensional
  fermi gases}},\ in\ \href {\doibase 10.1142/9789814667746_0001} {\emph
  {\bibinfo {booktitle} {Annual Review of Cold Atoms and Molecules}}},\
  Vol.~\bibinfo {volume} {3},\ \bibinfo {editor} {edited by\ \bibinfo {editor}
  {\bibfnamefont {K.~W.}\ \bibnamefont {Madison}}, \bibinfo {editor}
  {\bibfnamefont {K.}~\bibnamefont {Bongs}}, \bibinfo {editor} {\bibfnamefont
  {L.~D.}\ \bibnamefont {Carr}}, \bibinfo {editor} {\bibfnamefont {A.~M.}\
  \bibnamefont {Rey}}, \ and\ \bibinfo {editor} {\bibfnamefont
  {H.}~\bibnamefont {Zhai}}},\ Chap.~\bibinfo {chapter} {1}, pp.\ \bibinfo
  {pages} {1--75}\BibitemShut {NoStop}%
\bibitem [{\citenamefont {Petrov}\ \emph {et~al.}(2005)\citenamefont {Petrov},
  \citenamefont {Salomon},\ and\ \citenamefont {Shlyapnikov}}]{Petrov2005}%
  \BibitemOpen
  \bibfield  {author} {\bibinfo {author} {\bibfnamefont {D.~S.}\ \bibnamefont
  {Petrov}}, \bibinfo {author} {\bibfnamefont {C.}~\bibnamefont {Salomon}}, \
  and\ \bibinfo {author} {\bibfnamefont {G.~V.}\ \bibnamefont {Shlyapnikov}},\
  }\bibfield  {title} {\bibinfo {title} {\emph {Diatomic molecules in ultracold
  Fermi gases{\textemdash}novel composite bosons}},\ }\href {\doibase
  10.1088/0953-4075/38/9/014} {\bibfield  {journal} {\bibinfo  {journal}
  {Journal of Physics B: Atomic, Molecular and Optical Physics}\ }\textbf
  {\bibinfo {volume} {38}},\ \bibinfo {pages} {S645} (\bibinfo {year}
  {2005})}\BibitemShut {NoStop}%
\bibitem [{\citenamefont {Petrov}\ \emph {et~al.}(2003)\citenamefont {Petrov},
  \citenamefont {Baranov},\ and\ \citenamefont {Shlyapnikov}}]{Petrov2003}%
  \BibitemOpen
  \bibfield  {author} {\bibinfo {author} {\bibfnamefont {D.~S.}\ \bibnamefont
  {Petrov}}, \bibinfo {author} {\bibfnamefont {M.~A.}\ \bibnamefont {Baranov}},
  \ and\ \bibinfo {author} {\bibfnamefont {G.~V.}\ \bibnamefont
  {Shlyapnikov}},\ }\bibfield  {title} {\bibinfo {title} {\emph {Superfluid
  transition in quasi-two-dimensional Fermi gases}},\ }\href {\doibase
  10.1103/PhysRevA.67.031601} {\bibfield  {journal} {\bibinfo  {journal} {Phys.
  Rev. A}\ }\textbf {\bibinfo {volume} {67}},\ \bibinfo {pages} {031601}
  (\bibinfo {year} {2003})}\BibitemShut {NoStop}%
\bibitem [{\citenamefont {Bao}\ \emph {et~al.}(2019)\citenamefont {Bao},
  \citenamefont {Liu}, \citenamefont {Xue}, \citenamefont {Zheng},
  \citenamefont {Tao}, \citenamefont {Wang}, \citenamefont {Xia}, \citenamefont
  {Zhao}, \citenamefont {Kim}, \citenamefont {Yang}, \citenamefont {Li},
  \citenamefont {Wang}, \citenamefont {Wang}, \citenamefont {Wang},
  \citenamefont {MacDonald},\ and\ \citenamefont {Zhang}}]{Rypolariton}%
  \BibitemOpen
  \bibfield  {author} {\bibinfo {author} {\bibfnamefont {W.}~\bibnamefont
  {Bao}}, \bibinfo {author} {\bibfnamefont {X.}~\bibnamefont {Liu}}, \bibinfo
  {author} {\bibfnamefont {F.}~\bibnamefont {Xue}}, \bibinfo {author}
  {\bibfnamefont {F.}~\bibnamefont {Zheng}}, \bibinfo {author} {\bibfnamefont
  {R.}~\bibnamefont {Tao}}, \bibinfo {author} {\bibfnamefont {S.}~\bibnamefont
  {Wang}}, \bibinfo {author} {\bibfnamefont {Y.}~\bibnamefont {Xia}}, \bibinfo
  {author} {\bibfnamefont {M.}~\bibnamefont {Zhao}}, \bibinfo {author}
  {\bibfnamefont {J.}~\bibnamefont {Kim}}, \bibinfo {author} {\bibfnamefont
  {S.}~\bibnamefont {Yang}}, \bibinfo {author} {\bibfnamefont {Q.}~\bibnamefont
  {Li}}, \bibinfo {author} {\bibfnamefont {Y.}~\bibnamefont {Wang}}, \bibinfo
  {author} {\bibfnamefont {Y.}~\bibnamefont {Wang}}, \bibinfo {author}
  {\bibfnamefont {L.-W.}\ \bibnamefont {Wang}}, \bibinfo {author}
  {\bibfnamefont {A.~H.}\ \bibnamefont {MacDonald}}, \ and\ \bibinfo {author}
  {\bibfnamefont {X.}~\bibnamefont {Zhang}},\ }\bibfield  {title} {\bibinfo
  {title} {\emph {Observation of Rydberg exciton polaritons and their
  condensate in a perovskite cavity}},\ }\href {\doibase
  10.1073/pnas.1909948116} {\bibfield  {journal} {\bibinfo  {journal}
  {Proceedings of the National Academy of Sciences}\ }\textbf {\bibinfo
  {volume} {116}},\ \bibinfo {pages} {20274} (\bibinfo {year}
  {2019})}\BibitemShut {NoStop}%
\bibitem [{\citenamefont {Heck\"otter}\ \emph {et~al.}(2017)\citenamefont
  {Heck\"otter}, \citenamefont {Freitag}, \citenamefont {Fr\"ohlich},
  \citenamefont {A\ss{}mann}, \citenamefont {Bayer}, \citenamefont {Semina},\
  and\ \citenamefont {Glazov}}]{Heckotter2017}%
  \BibitemOpen
  \bibfield  {author} {\bibinfo {author} {\bibfnamefont {J.}~\bibnamefont
  {Heck\"otter}}, \bibinfo {author} {\bibfnamefont {M.}~\bibnamefont
  {Freitag}}, \bibinfo {author} {\bibfnamefont {D.}~\bibnamefont {Fr\"ohlich}},
  \bibinfo {author} {\bibfnamefont {M.}~\bibnamefont {A\ss{}mann}}, \bibinfo
  {author} {\bibfnamefont {M.}~\bibnamefont {Bayer}}, \bibinfo {author}
  {\bibfnamefont {M.~A.}\ \bibnamefont {Semina}}, \ and\ \bibinfo {author}
  {\bibfnamefont {M.~M.}\ \bibnamefont {Glazov}},\ }\bibfield  {title}
  {\bibinfo {title} {\emph {Scaling laws of Rydberg excitons}},\ }\href
  {\doibase 10.1103/PhysRevB.96.125142} {\bibfield  {journal} {\bibinfo
  {journal} {Phys. Rev. B}\ }\textbf {\bibinfo {volume} {96}},\ \bibinfo
  {pages} {125142} (\bibinfo {year} {2017})}\BibitemShut {NoStop}%
\bibitem [{Note1()}]{Note1}%
  \BibitemOpen
  \bibinfo {note} {The symmetry of $T({\protect \bf k},{\protect \bf k}';E)$ is
  straightforward to see by iterating Eq.~\protect \textup {\hbox
  {\mathsurround \z@ \protect \normalfont (\ignorespaces \ref {eq:Tmat}\unskip
  \@@italiccorr )}} once.}\BibitemShut {Stop}%
\bibitem [{2DG()}]{2DGreencorrect}%
  \BibitemOpen
  \href@noop {} {}\bibinfo {note} {Note that the corresponding equation (68) of
  Ref.~\cite{2DGreenfunction} had a small typo which we have corrected
  here.}\BibitemShut {Stop}%
\bibitem [{\citenamefont {Combescot}(2017)}]{3DCoulombTmatrix}%
  \BibitemOpen
  \bibfield  {author} {\bibinfo {author} {\bibfnamefont {R.}~\bibnamefont
  {Combescot}},\ }\bibfield  {title} {\bibinfo {title} {\emph {Three-Body
  Coulomb Problem}},\ }\href {\doibase 10.1103/PhysRevX.7.041035} {\bibfield
  {journal} {\bibinfo  {journal} {Phys. Rev. X}\ }\textbf {\bibinfo {volume}
  {7}},\ \bibinfo {pages} {041035} (\bibinfo {year} {2017})}\BibitemShut
  {NoStop}%
\bibitem [{\citenamefont {{Betbeder-Matibet, O.}}\ and\ \citenamefont
  {{Combescot, M.}}(2002)}]{Betbeder-Matibet2002}%
  \BibitemOpen
  \bibfield  {author} {\bibinfo {author} {\bibnamefont {{Betbeder-Matibet,
  O.}}}\ and\ \bibinfo {author} {\bibnamefont {{Combescot, M.}}},\ }\bibfield
  {title} {\bibinfo {title} {\emph {Commutation technique for interacting
  close-to-boson excitons}},\ }\href {\doibase 10.1140/epjb/e2002-00184-y}
  {\bibfield  {journal} {\bibinfo  {journal} {Eur. Phys. J. B}\ }\textbf
  {\bibinfo {volume} {27}},\ \bibinfo {pages} {505} (\bibinfo {year}
  {2002})}\BibitemShut {NoStop}%
\end{thebibliography}%

\end{document}